\documentclass[10pt,aps,prc,nofootinbib]{revtex4-2}
\usepackage{custom}
\usepackage{custom-3}

\begin{document}

\title{\bf Non-Gaussian fluctuations in relativistic hydrodynamics: 
\\[1ex]
\normalsize  Confluent equations for three-point correlations}

\date{\small\today}

\author{Xin An}\email{xin.an@ugent.be}
\affiliation{Department of Physics and Astronomy, Ghent University, 9000 Ghent, Belgium}

\author{G\"{o}k\c{c}e Ba\c{s}ar}\email{gbasar@unc.edu}
\affiliation{Department of Physics and Astronomy, University of North Carolina, Chapel Hill, North Carolina 27599, USA}

\author{Mikhail Stephanov}\email{misha@uic.edu}
\affiliation{Department of Physics, University of Illinois, Chicago, Illinois 60607, USA\medskip}

\begin{abstract}
   We derive deterministic equations for the evolution of {\em non-Gaussian} fluctuations in relativistic stochastic hydrodynamics. This is achieved by defining the {\em average} local Landau frame and corresponding fluctuating hydrodynamic variables. Fully nonlinear stochastic hydrodynamics is expressed in a unified multi-component matrix form. A novel relativistic formalism, also manifestly covariant under SO(3) rotations of the local spatial basis in the average local Landau frame, is introduced.  The equations describe correlators of {\em all} hydrodynamic variables, including fluctuating {\em velocity} (or momentum density) --- a nontrivial problem in {\em relativistic} hydrodynamics.  
\end{abstract}

\maketitle

\newpage 
\tableofcontents
\newpage

\section{Introduction}
\label{sec:intro}

Hydrodynamics is a remarkably universal theory \cite{Landau:2013fluid}. Hydrodynamic regime emerges naturally in practically any many-body system (classical or quantum) in conditions where the evolution of the system is slow enough to allow sufficient time for local thermal equilibration to be achieved. The primary focus of this paper is {\em relativistic} hydrodynamics, which has proven to be a remarkably successful and indispensable description of heavy-ion collisions \cite{Romatschke:2017ejr,Florkowski:2017olj}. Since hydrodynamic evolution is driven by the pressure of the fluid, heavy-ion collisions provide valuable information about the QCD equation of state, i.e., the dependence of pressure on temperature and (baryon) chemical potential.

Fluctuations in hydrodynamics are of primary interest in this context for two reasons. First, fluctuations in equilibrium are determined by derivatives of pressure (susceptibilities), which diverge at the QCD critical point. The knowledge of the location (and of the existence, in the first place) of such a point is fundamental to our understanding of the QCD phase diagram. Second, in heavy-ion collisions, fluctuations are measurable. In particular, a major goal of the Beam Energy Scan program at RHIC is to identify signatures of the QCD critical point using fluctuation measurements \cite{Stephanov:1998dy,Stephanov:1999zu,Aggarwal:2010cw,Bzdak:2019pkr,An:2021wof,Stephanov:2024xkn}.  

Most of the predictions of the expected signatures are based on the assumption of local thermodynamic {\em equilibrium} for the fluctuations~\cite{Bzdak:2019pkr}. However, since achieving equilibrium takes time, due to the conserved nature of the local thermodynamic variables, non-equilibrium effects, such as memory, could be important for describing fluctuations in heavy-ion collisions. The purpose of this paper is to obtain the system of equations governing the out-of-equilibrium evolution of fluctuations in hydrodynamics.

Fluctuations are an inherent feature of dissipative hydrodynamics. Dissipation (irreversibility) itself arises due to coarse-graining (integrating out) of the short-scale, fast degrees of freedom in an otherwise time-reversible (unitarily evolving) system. On long space and time scales, due to averaging involved in coarse-graining, the hydrodynamic equations appear deterministic. However, the randomness is prominent on shorter scales and is an essential counterpart of the dissipation. This randomness is described by noise terms in stochastic hydrodynamics. The equilibrium values of fluctuations are achieved via the balance between the fast local noise driving the fluctuations and the dissipation damping these fluctuations. If the system evolves, the balance shifts to different values, characterizing the evolved local equilibrium state. The evolution of the fluctuation measures, such as correlators,  are described by deterministic equations. These are the equations we derive in this paper.

Such deterministic equations have been derived earlier in limited contexts. In particular, in non-relativistic context in Ref.~\cite{Andreev:1978}.  In relativistic context, but limited to the Bjorken boost-invariance expansion or conformal equation of state in Refs.~\cite{Akamatsu:2017,Akamatsu:2018,Martinez:2018} and in the most general context, for arbitrary {\em relativistic} flow in Refs.~\cite{An:2019rhf,An:2019fdc}. In all the above cases only Gaussian fluctuations were considered.  

{\em Non-Gaussian} fluctuations are of primary interest for the QCD critical point search because the non-Gaussianity is more sensitive to the critical point than the magnitude of fluctuations \cite{Stephanov:2008qz}. In this context, the equations for non-Gaussian fluctuations have been derived, for the first time, in Ref.~\cite{An_2021}, however, only for the simplest hydrodynamics-like system --- nonlinear diffusion of a charge density. The flow was taken into account in Ref.~\cite{An_2023}. But, again, only the average flow was considered, while the {\em fluctuations} of the flow velocity were neglected.

In this paper we derive, for the first time, the deterministic evolution equations for the {\em non-Gaussian} fluctuations taking into account {\em all\/} hydrodynamic modes, including velocity fluctuations. In this paper, we limit our presentation to the case of three-point non-Gaussian correlators. However, the approach is general and, we believe, could be applied also to derive equations for four- and higher-point correlators. We leave these generalizations to future work.

One of the motivations for this work is the recent measurement by STAR collaboration of the fluctuations of proton multiplicities as a function of the center of mass energy in gold-gold collisions reported in Ref.~\cite{STAR:2025zdq}. In order to definitively interpret the experimental data, it is crucial to understand the effect of the non-equilibrium evolution of the fluctuations, including, and especially, of their non-Gaussianity. 

Interesting signals have been observed for (factorial) cumulants of the fluctuations of all orders $N\le4$ reported in Ref.~\cite{STAR:2025zdq}. While the critical point expectations for $N=4$ cumulants are more intricate, and thus harder to misattribute, the need for more data at larger $N$ potentially makes interesting signals in $N=2$ and $N=3$ cumulants more statistically significant in terms of signal-to-noise ratio. This makes focus on $N=3$ --- the lowest order of non-Gaussianity, especially warranted and timely. It also allows us to show the main ingredients of the novel formalism more clearly, paving the way for extensions to higher $N$.

\subsection{Hydrodynamics and scale hierarchy}
\label{sec:scales}

To set up the framework, we must emphasize that hydrodynamics is an effective theory in the sense that it relies on the separation of scales. The locality of hydrodynamic equations is achieved in so far as the spacial scale characterizing the inhomogeneity of the local thermodynamic conditions is much longer than the scale characterizing the fast and short-scale degrees of freedom being coarse-grained away to obtain local thermodynamic and hydrodynamic description. These short scales are often referred to as ``microscopic'' --- e.g., mean-free path in an underlying kinetic description, if it exists, or inverse temperature for a strongly coupled theory. 

We refer to the long hydrodynamic scale by the characteristic wavenumber $k$ --- inverse of the wave length of typical inhomogeneities in hydrodynamic quantities. This number must be small compared to the microscopic scales. Another scale which is important for hydrodynamics of fluctuations is the wavenumber, $q$, of the fluctuations. This also must be small compared to the microscopic scales in order for the fluctuations to be described by hydrodynamics.

The most interesting scale, as far as the fluctuations are concerned, is associated with the fluctuations which evolve at the rate comparable to the evolution of the fluid itself, i.e., $\mathcal O(c_s k)$, where $c_s$ is the speed of sound. Fluctuations relax diffusively, i.e., on the scale of $Dq^2$, where $D$ is the corresponding diffusion coefficient. This means that, in hydrodynamic regime, the corresponding wavenumber $q_*$ must be relatively large: $q_*\sim\sqrt k\gg k$ --- where we are using units of the microscopic scale in which both $q_*$ and $k$ are small. Fluctuations with wavenumbers much smaller than $q_*$ do not evolve and can be determined directly from initial conditions, while fluctuations with $q\gg q_*$ are always in equilibrium and their evolution is simply tracking the evolution of the local equilibrium conditions. Therefore, we focus on the scales $q\sim q_*$ and thus assume hierarchy, expressed in microscopic units, 
\begin{equation}\label{eq:kq-hierarchy}
    k\sim q^2\ll q\ll 1\,,
\end{equation}
which we shall rely on in this paper.

\subsection{The structure of the paper}

The paper is organized as follows. 

In Section~\ref{sec:confl-form} we introduce the average rest frame of the fluid (average Landau frame) and define the variables as the fluctuating densities in this non-fluctuating frame. This approach is different from the one that has been taken in the earlier literature \cite{An_2023}, where the frame itself was fluctuating.\footnote{It is also different from the approach known as ``density frame'' \cite{Armas:2020mpr,Basar:2024qxd,Bhambure:2024axa,Bhambure:2024gnf} in that the frame we use is not a fixed laboratory frame, but a frame dynamically determined by the fluid itself.} We write the most general, fully {\em nonlinear} stochastic equation for these variables in the standard hydrodynamic gradient expansion, i.e., truncating to the second order in spatial gradients (the lowest order in which dissipation occurs). We do not specify the coefficients, $ABCD$, in these equations and treat them as arbitrary, until later in the paper. 

In Section~\ref{sec:so3} we introduce the ``confluent'' formalism, see Ref.~\cite{An:2019rhf,An:2019fdc,An_2023}, which allows us to treat the fluid four-velocity as if it was uniform by parallel transporting (boosting) the quantities measured in different space-time points. The local spatial basis triad is chosen arbitrarily. This introduces additional SO(3) local gauge invariance, which we exploit by constructing equations using manifestly locally gauge covariant building blocks (derivatives and correlators). The resulting master equation \eqref{eq:u-A-nablaH-ao} is applicable to correlators of arbitrary order $N$.

We apply the master equation to obtain equations \eqref{eq:H-2pt} for $N=2$ correlators in Section~\ref{sec:2pt} and equations \eqref{eq:H-3pt} for $N=3$ correlators in Section~\ref{sec:3pt}. Equations \eqref{eq:H-2pt} are novel to the extent that they are using the novel formalism (averaged Landau frame) introduced here, even though equations for $N=2$ (i.e., Gaussian) correlators have been derived before in different approaches, e.g., in Ref.~\cite{An:2019fdc}. 
In contrast, Eq.~\eqref{eq:H-3pt} is the first of its kind, describing evolution of {\em non-Gaussian} fluctuations of {\em all} hydrodynamic variables, including fluctuating velocity.

In Section~\ref{sec:wigner} we apply Wigner transform, generalized to non-Gaussian correlators in Ref.~\cite{An_2021}, and obtain evolution equations for the generalized $N=2$ and $N=3$ Wigner functions. These functions describe correlations of wavelets of given wave-vectors $\bm q$ and allow us to take advantage of the separation of
scales $q\gg k$ discussed in Section~\ref{sec:scales}.
Eq.~\eqref{eq:W3-eqs} for $N=3$ Wigner function is one of the main results of the paper.

In Section~\ref{sec:equilibrium} we derive conditions on the so far arbitrary gradient expansion coefficients $ABCD$ imposed by the requirement that an equilibrium solution exists. In Section~\ref{sec:hydro-conserved} we show that the defining properties of hydrodynamics --- conservation laws and the second law of thermodynamics --- are responsible for fulfilling the equilibrium conditions derived in Section~\ref{sec:equilibrium}. 

In Section~\ref{sec:evolution-conserved} we show that, due to a special form of the gradient expansion coefficients $ABCD$ in a theory with degrees of freedom chosen to be fluctuating densities defined in Section~\ref{sec:ave-Landau}, i.e., the densities in the average Landau frame, the new equations we derived, such as Eq.~\eqref{eq:W3-eqs}, take simpler and more intuitive form.

Our equations are written in a matrix form, with hydrodynamic variables combined into a vector (one-dimensional array) and the coefficients into matrices.
In Section~\ref{sec:ABCD-hydro} we show how to determine the elements of these coefficient matrices from a given hydrodynamic input --- equation of state and transport coefficients.

In Section~\ref{sec:phonon} we show that the sector of the equations for the $N=2$ Wigner functions corresponding to the longitudinal velocity and pressure fluctuations (the sound sector) matches the kinetic equation for phonons. There is a very nontrivial and remarkable agreement which includes nontrivial inertial forces experienced by the phonons in an accelerating and rotating fluid, as well as other physically intuitive properties of the phonon gas. We view this agreement as a nontrivial check of the validity of our results. 

Section~\ref{sec:conclusion} contains a concise summary of the major ingredients of our approach and some concluding remarks.

To help the reader quickly find the definitions of various notations throughout the paper we provide a list in Appendix \ref{sec:notations}. 

\section{Confluent formalism}
\label{sec:confl-form}

\subsection{Stochastic hydrodynamic equations in averaged Landau frame}
\label{sec:ave-Landau}

We begin by defining the average local Landau frame as a frame in which the average momentum density of the fluid vanishes. In other words --- the average local rest frame of the fluid. Given the {\em fluctuating} energy-momentum tensor of the fluid $\breve T^\mu_{~\nu}$, the four-velocity $u$  of the average Landau frame is given by:\footnote{We use $g_{00}<0$ metric signature, so that $u^0=-u_0>0$, $u_\mu u^\mu=-1$.}
\begin{equation}\label{eq:ave-uT}
    -u_\mu \langle\bTmunu\rangle = \varepsilon u_\nu\,,
\end{equation}
i.e., it is the Landau frame for the average energy-momentum tensor $\langle\bTmunu\rangle$. Like $
\langle\bTmunu\rangle$, this four-velocity does not fluctuate.\footnote{
The actual (fluctuating) Landau frame $\breve u$ of the fluctuating fluid is given by 
\begin{equation}\label{eq:breve-uT}
    -\breve u_\mu\bTmunu = \breve\varepsilon \breve u_\mu\,.
\end{equation}
}
While the four-velocity of the frame does not fluctuate, the energy and momentum densities in this frame, $\teps$ and $\pi$, do. They are given by
\begin{equation}\label{eq:uT-e-pi}
    - u_\mu \bTmunu \equiv \teps
    u_\nu +  \pi_\nu\,.
\end{equation}
Note that four-vector $\pi$ has only 3 independent components, since $\pi\cdot u=0$ by definition.
One advantage of these variables is that they are {\em linearly} related to $\bTmunu$. As a result $\langle\teps\rangle=\varepsilon$ and $\langle\pi\rangle=0$. 

In addition to energy and momentum densities, a fluid carrying a conserved charge will also be characterized by that charge density, which is defined, similarly to Eq.~\eqref{eq:uT-e-pi}, as the time component of the fluctuating current $\breve J^\mu$ in the average Landau frame, i.e.,
\begin{equation}\label{eq:uJ}
    -u_\mu \breve J^\mu = \tn\,.
\end{equation}
Of course, one can use any set of variables, which are functions of these conserved densities. In this paper we shall refer to the variables $\{\pi,\teps,\tn\}$ as ``primary'' variables.

We shall denote the set of variables by $\tpsi_a$.  
For the primary variables the index $a$ runs through 3 components of the momentum $\pi$, the energy $\teps$ and whatever conserved charges involved. For a more general choice of variables the index $a$ labels the corresponding functions of primary variables.
We shall discuss more explicit form of hydrodynamics in terms of these variables later, as it is not needed yet for our discussion.
For now it would suffice to observe that, no matter what variables we choose, the hydrodynamic equations, which are essentially conservation equations, such as
\begin{equation}\label{eq:dT=0}
    \partial_\mu \bTmunu =0\,,\quad \partial_\mu \breve J^\mu=0\,,
\end{equation}
supplemented with constitutive equations for $\bTmunu$ and $\breve J^\mu$, truncated to the first order in derivatives (the lowest dissipative order) can be rewritten in the following general form\footnote{The order of $\nu\mu$ indices in $\tilde D^{\nu\mu}$ is important and is chosen here for future convenience.}:
\begin{equation}\label{eq:u.dpsi}
    (u\cdot\partial)\tpsi_a = A_{~a}^{\mu~b}\partial_\mu\tpsi_b + B_a^{(0)}
    +D_{~~a}^{\mu\nu~b}\partial_\mu\partial_\nu\tpsi_b 
    +\Dt_{~~a}^{\nu\mu~bc}(\partial_\mu\tpsi_b)(\partial_\nu\tpsi_c)
    +(\partial_\mu\eta^{\widehat c})C_{\widehat c a}^\mu+\eta^{\widehat c}\Ct_{\widehat ca}^{\mu~b}\partial_\mu\tpsi_b\,;
\end{equation}
\begin{equation}
  \label{eq:eta-eta}  \left\langle\eta^{\widehat c}(x_1)\eta^{\widehat d}(x_2)\right\rangle
    =2\delta^{\widehat c\widehat d}\delta^{(4)}(x_1-x_2)\,.
\end{equation}
The coefficients $ABCD\Ct\Dt$ are (nonlinear) functions of the stochastic variables $\tpsi$, but do {\em not} contain their derivatives, which are written out explicitly.
All derivatives on the RHS are {\em spatial in the average Landau frame}, i.e., omitting variable and noise indices $a$, $\hat c$, etc.,
\begin{equation}
u_\mu A^\mu=0\,;\quad u_\mu D^{\mu\nu}=0\,;\quad u_\mu\Dt^{\mu\nu}=0\,;\quad u_\mu C^\mu=0\,;\quad u_\mu\Ct^\mu=0\,.
\end{equation}
While we keep the derivatives of the {\em fluctuating} variables explicit in Eq.~\eqref{eq:u.dpsi}, the derivatives of the average (background) flow velocity $u$ are contained in $B_a^{(0)}$.
In Section~\ref{sec:hydro-conserved}  we shall demonstrate explicitly that the hydrodynamic equations do have this form and determine these coefficients in terms of familiar hydrodynamic parameters (equation of state and transport coefficients) in Section~\ref{sec:ABCD-hydro}. The local thermal/statistical noise $\eta^{\widehat c}$ is labeled by a different index $\widehat c$ which runs through the currents characterizing the transport of the conserved quantities. Typically, the noise $\eta^{\widehat c}$ is represented by a tensor of one rank higher than the variable in whose equation it appears. For example, the fluctuations of charge are driven by the noise in the current and the fluctuations in the momentum conservation equation are driven by the noise in the stress tensor. Correspondingly, the index $\widehat c$ runs through the components of all these noise tensors.

\subsection{``Confluentization''}
It would have been convenient if the average Landau frame velocity $u$ was constant throughout space and time. This is, of course, not the case in the relevant applications of hydrodynamics. In order to simplify dealing with the space-time dependence of the frame in which our variables are defined, we shall use the framework of confluent parallel transport which was introduced in Ref.~\cite{An:2019rhf}. 

We define a confluent derivative $\cfd$ as a covariant derivative such that the averaged velocity itself is confluently constant, i.e.,\footnote{In earlier papers, the confluent derivative and connection where denoted by $\bar\nabla$ and $\bar\omega$, respectively~\cite{An:2019rhf}.}
\begin{equation}\label{eq:barnabla-def}
    \cfd_\mu u^\nu\equiv\partial_\mu u^\nu + \ucon_{\mu\lambda}^\nu u^\lambda =0\,.
\end{equation}
The confluent connection $\ucon^\nu_{\mu\lambda}$ is the generator of the boost  
$\Lambda^\nu_{~\lambda}$ needed to change the average four-velocity $u(x')$ of the fluid at point $x'$ to $u(x)$: 
\begin{equation}\label{eq:Lambda-u}
    \Lambda (x,x') u(x')=u(x)\,,\quad\mbox{or}\quad \Lambda(x,x+h)=1+h\cdot\ucon\,,
\end{equation}
for infinitesimal four-vector $h$.
It is easy to write explicitly a confluent connection $\ucon$ satisfying Eq.~\eqref{eq:barnabla-def}:
\begin{equation}\label{eq:bar-omega}
    \ucon_{\mu\lambda}^\nu = u_\lambda\partial_\mu u^\nu - u^\nu\partial_\mu u_\lambda\,.
\end{equation}

Using 
\begin{equation}
    \cfd_\mu\pi_\nu =\partial_\mu\pi_\nu - \ucon_{\mu\nu}^\lambda\pi_\lambda\,,\quad \cfd\teps=\partial\teps\,,\quad \cfd\tn=\partial\tn\,,
\label{eq:nabla-pi}
\end{equation}
we can rewrite all derivatives in Eq.~\eqref{eq:u.dpsi} in terms of confluent derivatives:
\begin{equation}
    (u\cdot\cfd)\tpsi_a = A_a^{\mu~b}\cfd_\mu\tpsi_b + B_a
    +D_{~~a}^{\mu\nu~b}\cfd_\mu\cfd_\nu\tpsi_b 
    +\Dt_{~~a}^{\nu\mu~bc}(\cfd_\mu\tpsi_b)(\cfd_\nu\tpsi_c)
    +(\cfd_\mu\eta^{\widehat c})C_{\widehat c a}^\mu+\eta^{\widehat c}\Ct_{\widehat ca}^{\mu~b}(\cfd_\mu\tpsi_b)\,,
\label{eq:u.nablapsi-ABCD}
\end{equation}
where gradients of the average velocity $u$ are absorbed into the $B$ term: $B_a=B^{(0)}_a+A^{\mu~b}_{~a}\ucon_{\mu b}^c\tpsi_c$ 
\footnote{Since, according to Eq.~\eqref{eq:kq-hierarchy}, for the purposes of deriving the fluctuation equations, we count the background gradients (such as gradients of $u$) as being of order $k\sim q^2$ --- an order higher than the gradients $q$ of fluctuating fields, the coefficients $A,D,C$ remain the same (and do not contain background gradients) to the order in gradients we are working.}
and we defined $\ucon_{\mu b}^a\equiv0$ if $a$ or $b$ refer to scalar variables, such as $\teps$ or $\tn$. Note that Eq.~\eqref{eq:u.nablapsi-ABCD} is fully nonlinear in the sense that the coefficients $ABCD\Ct\Dt$ are, in general, nonlinear functions of $\tpsi$ as mentioned earlier. 

In this paper we focus on velocity fluctuations. These fluctuations are represented by three variables in the set $\tpsi_a$, such as three independent components of $\pi_\mu$ or three independent functions of these components (e.g., fluctuating velocity). In the local rest frame of the fluid these variables form a three-vector. To represent such vector variables we introduce, following Ref.~\cite{An:2019rhf}, a local Cartesian triad of four-vectors $e_a^\mu$, $a=1,2,3$, or $\bm e^\mu$, orthogonal to $u_\mu$: 
\begin{equation}\label{eq:e.u}
    \bm e^\mu u_\mu\equiv \bm e\cdot u=0\,;\quad e^\mu_a e_\mu^b\equiv e_a\cdot e^b=\delta_a^b\,;\quad \bm e^\mu\cdot \bm e_\nu= \delta^\mu_\nu+u^\mu u_\nu\equiv\Delta^\mu_\nu\,.
\end{equation}

Similarly to Eq.~\eqref{eq:barnabla-def}, we introduce a confluent connection $\mathring\omega$ such that the local Cartesian triad is also confluently constant:
\begin{equation}
    \cfd_\mu e_a^\nu \equiv
    \partial_\mu e_a^\nu + \ucon_{\mu\lambda}^\nu e_a^\lambda - \mathring\omega_{\mu a}^b e^\nu_b=0\,.
\label{eq:nabla-e}\end{equation}
Solving this equation for $\mathring\omega$ we find:
\begin{equation}\label{eq:ring-o}
    \mathring\omega_{\mu a}^b
    = e^b_\nu\partial_\mu e^\nu_a 
    + e^b_\nu\ucon_{\mu\lambda}^{\nu}e_a^\lambda = e^b_\nu\partial_\mu e^\nu_a\,,
\end{equation}
where to obtain the last equation we used Eqs.~\eqref{eq:bar-omega} and~\eqref{eq:e.u}.
Since index $a$, in our earlier convention, runs over the whole set of hydrodynamic variables, in order to reuse the same notation for the index the summation over which runs only through the velocity/momentum variables, i.e., $a=1,2,3$, we shall extend the definition of $e_a^\mu$ so that for all $a\neq 1,2$ or $3$: $e_a^\mu=0$.

\subsection{Fluctuations and Taylor expansion}

Starting from the fully nonlinear confluent and covariant equation \eqref{eq:u.nablapsi-ABCD} we now expand in powers of fluctuations $\phi$ around the averaged values of variables:
\begin{equation}\label{eq:phi}
    \phi\equiv\tilde\psi-\psi\,,\quad\mbox{where}\quad \psi\equiv\langle\tpsi\rangle\,.
\end{equation}
In this work, in order to introduce the new approach, we focus on the lowest order non-Gaussian correlators, i.e., we truncate the expansion in Eq.~\eqref{eq:u.nablapsi-ABCD} at the lowest nonlinear, i.e., bilinear, order:
\begin{multline}\label{eq:u.phi-ABCD}
    u\cdot\cfd\phi_a
  =(B+A^{\mu}\cfd_\mu + 
     D^{\mu\nu}\cfd_\mu\cfd_\nu)_{a}^{~b}\phi_{b}+C^{\mu}_{~\widehat ca}\cfd_\mu\eta^{\widehat c}
    \\
    +\subtr{\left( B_{(3)}\mathbb 1
    +A_{(3)}^{\mu}\cfd_\mu\otimes1
    +
    D_{(3)}^{\mu\nu}\cfd_\mu\cfd_\nu\otimes1
 + {\Dt}_{(3)}^{\nu\mu}\cfd_\mu\otimes\cfd_\nu\right) ^{~bc}_{a~~}
 \args{\phi_{b},\phi_{c}}}
\\
 + \left(C_{(3)}^{\mu}1\otimes\cfd_\mu + {\Ct}_{(3)}^\mu \cfd_\mu\otimes1
\right)_{\widehat c a}^{~~b}
\args{\phi_{b},\eta^{\widehat c}}\,,
\end{multline}
where $\subtr{\phi_b\phi_c}\equiv \phi_b\phi_c-\langle\phi_b\phi_c\rangle$.
To reduce index clutter we introduced a shorthand: $(B_{(3)}+\dots)_a^{~bc}\equiv (B_a^{~bc}+\dots)$ and a similar shorthand $A^\mu_{(3)}$, etc.
To unambiguously represent bilinear (or multilinear) operators acting on two (or more) variables we introduced a notation: 
\begin{equation}\label{eq:fgxy}
    f\otimes g\args{x,y}\equiv f(x)g(y)\,;
    \quad
    \mathbb1\args{x,y}\equiv1\otimes1\args{x,y}= xy\,.
\end{equation}

The coefficients in Eq.~\eqref{eq:u.phi-ABCD} are obtained by Taylor expansion of coefficients in the fully nonlinear equation \eqref{eq:u.nablapsi-ABCD}. For example, 
\begin{equation}\label{eq:A3A'}
A_{~a}^{\mu~bc}=\frac{\partial A^{\mu~b}_{~a}}{\partial\tpsi_c}\Big|_{\tpsi=\psi}\equiv 
A_{~a}^{\mu~b,c}\,.
\end{equation}
The coefficient $B$, in addition to Taylor expansion of $B^{(1)}$, also receives contribution from the (confluent) background gradients $\cfd\psi$:
\begin{equation}\label{eq:B2B'}
B_a^{~b}=\left( B_a + 
A_{~a}^{\mu~d}\cfd_\mu\psi_d\right)^{,b}\,.
\end{equation}
Similarly to Eq.~\eqref{eq:A3A'}
\begin{equation}\label{eq:BDC'}
    B_a^{~bc}=\frac12 B_a^{~b,c}\,;
    \quad
    D_a^{~bc}=D_a^{~b,c}\,;
    \quad
    C_a^{~bc}=C_a^{~b,c}\,.
\end{equation}
The factor $1/2!$ appears because $B$ is already a derivative and thus $B_{(3)}$ is a  {\em second} derivative, according to Eq.~\eqref{eq:B2B'}.

\section{SO(3) covariant formalism for correlation functions}
\label{sec:so3}

\subsection{Covariant confluent correlator}
\label{sec:cc-correlator}

Note that the connection $\mathring\omega$ for confluent transport of local basis triad depends on the arbitrary local choice of such a triad. Since physics cannot depend on this arbitrary choice, the description we introduce possesses a local SO(3) gauge invariance. We shall exploit this invariance by ensuring that all our equations are explicitly SO(3) gauge invariant. This is already evident in the form of the stochastic equation \eqref{eq:u.nablapsi-ABCD}. In this section, we shall develop novel formalism which extends this desirable property to correlation functions, their derivatives, and thus their evolution equations.

Our first step is to write down SO(3) covariant generalization of equations which express a (local rest frame) time derivative of a correlation function in terms of a correlation function of the time derivative of the field. We can then apply equation of motion for the field, Eq.~\eqref{eq:u.nablapsi-ABCD}, to express the result in terms of spatial derivatives. 

Since three hydrodynamic variables, and, thus, their fluctuations $\phi_a$, $a=1,2,3$, are components of an SO(3) vector,  the correlation function 
\begin{equation}
\rawG_{a_1\dots a_N}(x_1,\dots,x_N)\equiv   \langle\phi_{a_1}(x_1)\dots\phi_{a_N}(x_N)\rangle 
\label{eq:rawG}\end{equation}
transforms under
space-time dependent rotation of the local basis 
\begin{equation}\label{eq:eTe}
    e_a(x)\to  T_a^{~b}(x)e_b(x) 
\end{equation}
as follows:
\begin{equation}\label{eq:H-TTH}
\rawG_{a_1\dots a_N}(x_1,\dots,x_N)\to T_{a_1}^{~b_1}(x_1)\dots T_{a_N}^{~b_N}(x_N)\rawG_{b_1\dots b_N}(x_1,\dots,x_N)\,.
\end{equation}
What we want instead is a correlation function which is a {\em tensor} transforming locally with the transformation of the basis $T(x)$ at the midpoint:
\begin{equation}\label{eq:x}
    x\equiv\frac{1}{N}\sum_{i=1}^N {x_i} \,. 
\end{equation}
We, therefore, define {\em confluent} correlators
\begin{equation}
  \label{eq:H}
  \barG_{a_1\dots a_N}(x_1,\dots,x_N)
  \equiv \left\langle\prod_{i=1}^N R(x,
  x_i)^{~b_i}_{a_i}\phi(x_i)_{b_i}\right\rangle \equiv \left\langle\prod_{i=1}^N \left[R(x,
  x_i)\phi(x_i)\right]_{a_i}\right\rangle 
  \equiv \left\langle\prod_{i=1}^N \left[R\phi(x_i)\right]_{a_i}\right\rangle\,,
\end{equation}
where the last two expressions are shorthanded versions reducing index clutter.
We introduced $R(x,x_i)$ ---  the SO(3) connection matrix between points $x_i$ and $x$: 
\begin{equation}\label{eq:ReLe}
    R(x,x_i)_a^{~b}\equiv [\Lambda(x_i,x)e(x)_a]
\cdot    e(x_i)^b\,.
\end{equation}
The definition in Eq.~\eqref{eq:H} ensures that under a rotation of the local triad given
by Eq.~\eqref{eq:eTe}, which leads to 
\begin{equation}
    R(x,x')\to T(x)R(x,x')T(x')^{-1}\,,
\end{equation}
  the correlator $\barG$ transforms covariantly as
\begin{equation}
  \label{eq:H-TH}
  \barG_{a_1\dots a_N}(x_1,\dots x_N) \to 
  \left(\prod_{i=1}^N T_{a_i}^{~b_i}(x)\right)
   \barG_{b_1\dots b_N}(x_1,\dots x_N)\,,
\end{equation}
unlike the ``raw'' correlator $\rawG$ in Eq.~\eqref{eq:H-TTH}.

In Eq.~\eqref{eq:ReLe}, the boosted triad $e_a(x)$ is orthogonal to $u(x_i)$, i.e., $u(x_i)\cdot[\Lambda(x_i,x)e_a(x)]=0$ due to Eqs.~\eqref{eq:Lambda-u} and~\eqref{eq:e.u}. Since $e_b(x_i)$ is also a triad orthogonal to $u(x_i)$, matrix $R$ is an orthogonal rotation matrix.\footnote{While $\Lambda(x_i,x)$ is the boost aligning $u(x)$ with $u(x_i)$,
$R(x_i,x)$ is the spatial rotation aligning the basis triad $\bm e(x)$, after the same boost $\Lambda(x_i,x)$, with the basis triad $\bm e(x_i)$, i.e., $e(x_i)_b=R(x_i,x)_b^{~a}[\Lambda(x_i,x)e(x)_a]$.
} The connection $\mathring\omega$ is a generator for this rotation. This is easy to check by substituting, for $h\to0$,
\begin{equation}\label{eq:R-omega}
    R(x,x+h)=1-h\cdot\mathring\omega
\end{equation}
together with Eq.~\eqref{eq:Lambda-u} into Eq.~\eqref{eq:ReLe} and comparing to Eq.~\eqref{eq:ring-o}.

We shall consider {\em equal time} correlators. The time is measured in the rest frame of the fluid at midpoint~$x$. This means, in Eq.~(\ref{eq:H}), vectors $x_i$ can be expressed as 
\begin{equation}
    x_i=x+y_i\quad \mbox{with}\quad y_i=\bm e(x)\cdot\bm y_i\,.
\label{eq:xxy}
\end{equation}
Note that, by definition, the vectors $y_i$ are linearly {\em dependent}:
$\sum_{i=1}^Ny_i=0 $.
It will be convenient at certain times to express the arguments of the confluent correlator in terms of $x$ and a set of three-vectors $\bm y_1,\dots,\bm y_N\equiv\{\bm y_i\}_{i=1}^N$:
\begin{equation}
\barG_{a_1\dots a_N}(x_1,\dots,x_N)\equiv\barG_{a_1\dots a_N}\left(x;\bm y_1,\dots,\bm y_N 
\right)\,.
\end{equation}
Because $\bm y$ are coordinates in the local basis $\bm e(x)$, they change under the local rotation of the basis, i.e.,
\begin{equation}
  \label{eq:H-TH-y}
  \barG_{a_1\dots a_N}(x;\{\bm y_i\}_{i=1}^N)\to \left(\prod_{i=1}^N T_{a_i}^{~b_i}(x)\right)
  \barG_{b_1\dots b_N}(x;\{ T(x)^{-1}\bm y_i\}_{i=1}^N)\,.
\end{equation}

Later in the paper we shall find convenient to elevate the space-time arguments $x_i$ or $\bm y_i$ to the superscript positions, which will emphasize their correspondence to field indices $a_i$ and save some space, i.e., for future reference,
\begin{equation}
\barG_{a_1\dots a_N}^{(x_1,\dots,x_N)}\equiv\barG_{a_1\dots a_N}^{(\bm y_1,\dots,\bm y_N)}(x)\equiv \barG_{a_1\dots a_N}(x_1,\dots,x_N) \,.
\label{eq:G-sup}
\end{equation}
We shall also find it convenient to suppress the indices and/or space-time arguments where this does not create ambiguity, so that,
\begin{equation}
\barG_{a_1\dots a_N}(x_1,\dots,x_N)
\equiv \barG_{a_1\dots a_N}\equiv \barG_{(N)}\,.
\label{eq:GN}
\end{equation}

\subsection{Covariant confluent derivatives of a correlator}
\label{sec:cc-derivs}

To make the following calculations more manageable let us introduce two linear operators  $Z_h$ and $\Delta_\hstep$ which act on an arbitrary function $f$ as follows:
\begin{equation}
  \label{eq:Deps-def}
  Z_\hstep f(\hstep)\equiv f(0)\,,\quad \Delta_\hstep f(\hstep) \equiv f(\hstep)-f(0)\,,
\end{equation}
where $\hstep$ could be a multicomponent quantity, e.g., a vector. While $Z_\hstep$ simply evaluates $f$ at $\hstep=0$, $\Delta_h$ gives the {\em finite} change of the function due to the shift of its argument from $0$ to $\hstep$ and is convenient for constructing derivatives, of course.
Though, in practice, we shall consider $\hstep$ to be infinitesimally small, Eq.~\eqref{eq:Deps-def} defines $\Delta_\hstep$ for arbitrary {\em finite} $\hstep$. Note that $Z_h+\Delta_h=1$ is an identity operator.
The change of a function of several arguments due to simultaneous change of the arguments by the same amount $h$ can be represented as follows:
\begin{multline}
 \label{eq:f-rule-N}
  \Delta_\hstep f({\hstep,\dots,\hstep})
=\left(\prod_{i=1}^N(Z_{\hstep_i}+\Delta_{\hstep_i})\right)f(\hstep_1,\dots,\hstep_N)\Big|_{\hstep_1=\dots=\hstep_N=\hstep}-f(0,\dots,0)
\\=\left[\left(
N\Delta_{\hstep_1}Z_{\hstep_2}\dots Z_{\hstep_N}+\binom{N}{2}\Delta_{\hstep_1}\Delta_{\hstep_2}Z_{\hstep_3}\dots Z_{\hstep_N}+\dots
\right)f(h_1,\dots,h_N)
\right]_\oN\Bigg|_{h_1=\dots=h_N=h} 
\,,
\end{multline}
where we expanded the product of binomials $Z_{\hstep_i}+\Delta_{\hstep_i}$ and showed only the terms of order $h$ and $h^2$.
The notation $[\dots]_\oN$ signifies the average over permutations of the indices $1\dots N$.
Although, for infinitesimal~$\hstep$, the
$\mathcal O(\Delta_\hstep^2)$ term is negligible for regular functions $f$, this term will
be important for the calculus of stochastic functions.

Using this notation we can write, for example, the confluent derivative of a field fluctuation variable $\phi$:
\begin{equation}
  \label{eq:Dphi-Delta}
  \Dx\cdot\cfd\phi(x) \equiv \Delta_{\Dx} [R(x,x')\phi(x')]\,,
  \quad\mbox{with}\quad x'\equiv x+\Dx\,,
\end{equation}
where we used matrix notations $[R\phi]_a\equiv R_a^{~b}\phi_b$ to suppress index clutter. More explicitly,
\begin{equation}\label{eq:Dphi-omega}
    \cfd\phi = \partial\phi - \mathring\omega\phi\,,
\end{equation}
where we used Eq.~\eqref{eq:f-rule-N} and Eq.~\eqref{eq:R-omega}, which can be also written as $\Dx\cdot\mathring \omega\equiv -\Delta_{\Dx}R(x,x+\Dx)$
(here and throughout we keep only the leading terms in the $\Dx\to0$ limit).
In what follows, we shall not use Eq.~\eqref{eq:Dphi-omega}, but instead operate directly with the definition in Eq.~\eqref{eq:Dphi-Delta}.

More generally, for an $N$-point function,  we define the confluent derivative with respect to the midpoint~$x$ via
\begin{multline}
  \label{eq:epsnabla}
  \Dx\cdot\cfd \barG_{a_1\dots a_N} \left(
  x_1,\dots,x_N
  \right)
  \equiv \Delta_{\Dx}\left[ \left(\prod_{i=1}^N R(x,x')_{a_i}^{~b_i}\right)
  \barG_{b_1\dots b_N}\left(
  x_1',\dots,x_N'
  \right)\right]
  \\=
  \Delta_{\Dx}\left\langle
    \prod_{i=1}^N [R(x,x')R(x',x_i')\phi(x_i')]_{a_i}
  \right\rangle\,,
\end{multline}
where
\begin{equation}\label{eq:x'}
x'\equiv x+\Dx\,,\quad x_i'\equiv x' + \bm e(x')\cdot R(x',x)\,\bm y_i=
x+\Dx + \bm e(x+\Dx)\cdot R(x+\Dx,x)\,\bm y_i\,,
\end{equation}
and we omit indices, where this does not create ambiguity, to minimize unnecessary clutter.
This definition ensures that under a rotation of the local triad the confluent derivative $\cfd \barG$ transforms in
the same way as the confluent correlator $\barG$ itself, i.e., as in Eqs.~\eqref{eq:H-TH} and~\eqref{eq:H-TH-y}.

It is convenient to use the following (transitive) property of $R$:
\begin{equation}
  \label{eq:RR-R}
  R(x,x')R(x',x_i') = R(x,x_i') + \mathcal O (\mathring\omega^2 y_i )\Dx\,,
\end{equation}
which follows from Eq.~\eqref{eq:R-omega}.
 The residual term is subleading, since $\mathring\omega=\mathcal
O(k)$, and is negligible within the hydrodynamic gradient truncation order we consider. Thus, within the necessary precision,
\begin{equation}
  \label{eq:epsnabla-R}
  \Dx\cdot\cfd \barG_{a_1\dots a_N} \left(
  x_1,\dots,x_N
  \right)
 =
  \Delta_{\Dx}\left\langle
    \prod_{i=1}^N [R(x,x_i')\phi(x_i')]_{a_i}\right\rangle\,.
\end{equation}

Now we can apply the rule in Eq.~(\ref{eq:f-rule-N}) obeyed by $\Delta_{\Dx}$ to express this as
\begin{multline}
\label{eq:epsnabla-sumnablax1H}
  \Dx\cdot\cfd \barG_{a_1\dots a_N}
  x_1,\dots,x_N
  )
 =
  \Bigg[N\left\langle \Delta_{\Dx}[R(x,x_1')\phi(x_1')]_{a_1}
    \prod_{i=2}^N [R(x,x_i)\phi(x_i)]_{a_i}
  \right\rangle\\
  +\frac{1}{2}N(N-1)\left\langle \Delta_{\Dx}[R(x,x_1')\phi(x_1')]_{a_1}\Delta_{\Dx}[R(x,x_2')\phi(x_2')]_{a_2}
    \prod_{i=3}^N [R(x,x_i)\phi(x_i)]_{a_i}
  \right\rangle+\mathcal O(\Delta_\Dx^3)\Bigg]_{\overline{1\dots N}}\,.
\end{multline}
The second, $\mathcal O(\Delta_\Dx^2)$ term is important for stochastic calculus.
It is convenient to define
\begin{equation}
  \label{eq:nablax1H-nablaphi}
  \Dx\cdot\dxone  \barG_{a_1\dots a_N}(x_1,\dots,x_N) \equiv
  \Delta_{\Dx} \left\langle [R(x,x_1+\Dx)\phi(x_1+\Dx)]_{a_1}
    \prod_{i=2}^N [R(x,x_i)\phi(x_i)]_{a_i}
  \right\rangle\,,
\end{equation}
i.e., using Eq.~\eqref{eq:Dphi-Delta} and a property akin to Eq.~\eqref{eq:RR-R},
\begin{equation}
  \label{eq:udx1H}
 \dxone  \barG_{a_1\dots a_N} (x_1,\dots,x_N) =
  \left\langle
[R(x,x_1)\cfd\phi(x_1)]_{a_1}\prod_{i=2}^N[R(x,x_i)\phi(x_i)]_{a_i}
  \right\rangle\,,
\end{equation}
which is important because it is covariant (i.e., transforms as $\barG$ itself) and contains $\cfd\phi(x_1)$ which we can manipulate using equations of motion. 

We enclose index $1$ in square brackets in the notation $\dxone$ to emphasize that under an index permutation, such as $\oN$ in Eq.~\eqref{eq:epsnabla-sumnablax1H}, this derivative always refers to the first argument of the correlator, no matter what name (which can change under permutation) this argument has.

Similarly, we define
\begin{equation}
     \dxone\dxtwo  \barG_{a_1\dots a_N} (x_1,\dots,x_N) \equiv
  \left\langle
[R(x,x_1)\cfd\phi(x_1)]_{a_1}[R(x,x_2)\cfd\phi(x_2)]_{a_2}\prod_{i=3}^N[R(x,x_i)\phi(x_i)]_{a_i}
  \right\rangle\,.
\end{equation}
Again, $\dxone$ and $\dxtwo$ refer to the first and the second argument in the correlator, respectively, no matter what symbol we use to denote this variable, $x_1$ and $x_2$ in the case at hand.

While these definitions appear very similar to the terms on the RHS\ of Eq.~\eqref{eq:epsnabla-sumnablax1H}
there is
a subtlety here in that, according to Eq.~(\ref{eq:x'}), $x_1'$ in Eq.~(\ref{eq:epsnabla-sumnablax1H}) is not exactly $x_1+\Dx$, but
$x_1+\Dx+\Delta_{\Dx} \bm e(x+\Dx)\cdot R(x+\Dx,x)\bm y_1$. Since $\bm e$ is confluently constant, i.e., $\Delta_{\Dx}\Lambda(x,x')\bm e(x')
R(x',x)=0$, we can write (using Eq.~(\ref{eq:f-rule-N}) again)
$\Delta_{\Dx}\bm e(x')R(x',x)=-\Delta_{\Dx}\Lambda(x,x')\bm
e(x) = -\Dx\cdot \ucon\bm e$. This additional term is smaller by a factor of $\mathcal O(ky_1)$ because $\ucon=\mathcal O(k)$ and since the $\mathcal O(\Delta_\Dx^2)$ term in Eq.~\eqref{eq:epsnabla-sumnablax1H} already has two gradients, this additional term can be neglected there. However, this additional contribution is important in the $\mathcal O(\Delta_\Dx)$ term in Eq.~\eqref{eq:epsnabla-sumnablax1H} and we obtain:
\begin{equation}
  \label{eq:nablaH-sum-nablax1H}
  \Dx\cdot \cfd \barG _{(N)}=N\left[h\cdot\dxone\barGN
    -  \Dx^\nu\ucon_{\nu\lambda}^\mu y_1^\lambda\dxone_\mu\barGN+
    \frac{N-1}2\Dx\cdot\dxone\,\Dx\cdot\dxtwo\barGN
    \right]_\oN
\end{equation}
where we introduced $y^\lambda\equiv\bm e^\lambda\cdot\bm y_1$. We are suppressing indices and space-time arguments using notation in Eq.~\eqref{eq:GN}.
While the last term appears to be $\mathcal O(\Dx^2)$, due to the singularity of the noise correlator in Eq.~\eqref{eq:eta-eta} when $h\to0$, this term will give rise to an important contribution of the same order, $\hstep$, as the first two terms.

According to its definition in Eq.~\eqref{eq:nablax1H-nablaphi}, $\dxone\barGN$ is a {\em correlator of a derivative} of the fluctuation field $\phi$.
As such, it is an intermediate object which we would like to express in terms of the correlator $\barGN$ and its derivatives using equations of motion for the field $\phi$.
For that purpose, we shall introduce another derivative of a correlator (which is important because its Wigner
transform is simple):
\begin{equation}
  \label{eq:dy1}
  \Dby\cdot\bdyone \barGN(x;\{\bm y_i\}_{i=1}^N)  =
  \Delta_{\Dby}\barGN\left(x;\{\bm y_i+\Dby_i\}_{i=1}^N\right)
  \end{equation}
  where the displacements are spatial in the fluid's average rest frame at midpoint and
  \begin{equation}
    \label{eq:Dyi}
    \Dby_i=\left(\delta_{i1}-\frac1N\right)\Dby\,.
  \end{equation}
The displacements $\Dby_i$ do not move the midpoint, i.e., $\sum_{i=1}^N\Dby_i=0$. To reflect this property, we shall refer to this derivative as a {\em balanced} derivative.
As a result, unlike standard partial derivatives $(\partial/\partial\bm y_i)$, the balanced derivatives $\bdyi$  are not independent and obey \begin{equation}
\sum_{i=1}^N\bdyi=0\,.
\end{equation}
More importantly,
\begin{equation}\label{eq:d-sum-y}
\bdyone\sum_{i=1}^N\bm y_i=0\,.    
\end{equation}
This makes the Wigner transform of such derivatives simple
since it is not affected by $\delta^{(3)}\left(\sum_{i=1}^N\bm y_i\right)$ factor which distinguishes Wigner transform from a standard Fourier transform (see below, Eq.~\eqref{eq:WG}).

By definition, the balanced derivative $\bdyone $ is spatial in the rest frame of the fluid at $x$. The balanced derivative can be easily expressed in terms of
$\dxone $. To achieve that we apply the property in Eq.~(\ref{eq:f-rule-N}) after separating $\Dby_i$
into $i$-independent shift $-\Dby/N$ of all $y_i$'s and the shift
of only $\bm y_1$ by $\Dby$.\footnote{The $\mathcal O(\Delta_{\Dby}^2)$ terms in Eq.~\eqref{eq:f-rule-N} do not contribute here even for stochastic calculus because {\em time-like} derivatives are not involved.}
We find
\begin{equation}
  \label{eq:y1-x1-x1}
  \bdyone \barGN= \bm e\cdot\left(\dxone \barGN-\left[\dxone 
    \barGN\right]_{\overline{1\dots N}} \right) \,.
\end{equation}
In practice, we shall use this relation to express $\bm e\cdot\dxone$ in terms of $\bdyone$. In particular,
\begin{equation}
  \label{eq:x1-x-y1}
  \bm e\cdot \dxone\barGN
  =   \bdyone \barGN
+   \frac1N\bm e\cdot\cfd \barGN
    +\left[(\bm e\cdot\ucon)_{~\lambda}^\mu y_1^\lambda
     \dxone_{\mu}\barGN\right]_{\overline{1\dots N}}
\,,
\end{equation}
where we used Eq.~\eqref{eq:nablaH-sum-nablax1H}, neglecting the last term because the displacement $h$ in this case is spatial and the singularity due to stochastic calculus does not arise.
Since the last two terms in Eq.~\eqref{eq:x1-x-y1} are of order $\mathcal O(k)$, to leading order in gradient expansion, i.e., $\mathcal O(q)$,
the spatial
components of $\dxone$ are the same as those of $\bdyone$:
\begin{equation}
  \label{eq:x1-y1}
  \bm e\cdot \dxone \barGN = \bdyone \barGN + \mathcal O(k)\,.
\end{equation}

We shall also find useful a four-vector representation of the balanced derivative:
\begin{equation}\label{eq:dy-bdy}
    \dyone{\lambda}\equiv\bm e_\lambda\cdot\bdyone\,,
\end{equation}
which is orthogonal to $u(x)$, $u\cdot\dyone{}=0$, unlike $\dxone$.

\subsection{Master equation}
\label{sec:master}

Our main goal is to derive the evolution equation for $\barGN$ which is an expression for the time derivative of $\barGN$ in the local rest frame of the fluid in terms of spatial derivatives in the frame. To define the corresponding confluent derivatives, we shall use displacement $h$ in the direction of the average flow velocity~$u(x)$: \begin{equation}\label{eq:Dxt}
    \Dx=\Dxt u(x), 
\end{equation}
where $\Dxt$ is an infinitesimal scalar.
The equations of motion given in  Eq.~\eqref{eq:u.phi-ABCD} can be used to convert time derivative $u(x_1)\cdot\cfd\phi(x_1)$ into spatial derivatives. In doing so, we need to take into account that the velocity $u(x_1)$ is different from the velocity $u(x)$, i.e.,
(up to higher order in gradients)
\begin{equation}
    u^\mu(x_1)-u(x)^\mu=(y_1\cdot\partial) u^\mu =
    -y_1^\lambda\ucon_{\lambda\nu}^\mu u^\nu\,,
\end{equation}
where we omit the argument $(x)$ when expressing terms where the difference between using $x$ or $x_1$ is negligible (higher order in gradients). Therefore, according to Eq.~\eqref{eq:udx1H},
\begin{equation}
   \label{eq:u.nabla-x1} u\cdot\dxone\barGN = (u\cdot\der)^\xone\barGN + y_1^\lambda\ucon_{\lambda\nu}^\mu u^\nu\dxone_\mu\barGN\,,
\end{equation}
where $(u\cdot\der)^\xone \equiv u(x_1)\cdot\dxone$. Substituting this and $\hstep=\bar\hstep u$ into Eq.~\eqref{eq:nablaH-sum-nablax1H} we obtain an equation which will be useful in deriving equations of motion for the correlator:
\begin{equation}
  \label{eq:nablaH-sum-nablax1H-torsion}
   u\cdot \cfd \barG _{(N)}=N\left[(u\cdot\der)^\xone\barGN
    +  y_1^\lambda \tors_{\lambda\nu}^\mu u^\nu\dxone_\mu\barGN+
    \frac{N-1}2\Dxt (u\cdot\der)^\xone (u\cdot\der)^\xtwo\barGN
    \right]_\oN 
\end{equation}
where  we introduced
\begin{equation}\label{eq:torsion}
    \tors^\mu_{\lambda\nu}
\equiv\ucon^\mu_{\lambda\nu}-\ucon^\mu_{\nu\lambda}\,,
\end{equation}
which we refer to as confluent torsion.

Because the leading term in the equation \eqref{eq:u.nablapsi-ABCD} for $u\cdot\cfd\tpsi$  is $A\cdot\cfd\tpsi$, it is convenient to write equations which involve correlators of $(u-A)\cdot\cfd\tpsi$. Such correlators will appear when we subtract from $u\cdot\cfd\barGN$ in Eq.~\eqref{eq:nablaH-sum-nablax1H-torsion} 
the following object
\begin{equation}\label{eq:A.nablaH}
    \left[A^{\mu~b}_{~a_1}\cfd_\mu \barG_{ba_2\dots a_N}
   \right]_\oN\equiv
    A\cdot\cfd \barG_{a_1\dots a_N}\,
\end{equation}
where $A$ is the vector-valued matrix of coefficients in the equation of motion Eq.~\eqref{eq:u.phi-ABCD}.
We would like to express this object in terms of a covariant correlator which involves $A\cdot\cfd\phi$, similar to how we expressed $u\cdot\cfd\barG$ in terms of the correlator 
$u(x_1)\cdot\dxone\barGN$
which involves $u\cdot\cfd\phi$ in Eq.~\eqref{eq:nablaH-sum-nablax1H-torsion}. We define the corresponding correlator as
    \begin{equation}
  \label{eq:Abdx1H}
 (A\cdot\der)^\xone  \barG_{a_1\dots a_N}  \equiv
  \left\langle
[R(x,x_1)A(x_1)\cdot\cfd\phi(x_1)]_{a_1}\prod_{i=2}^N[R(x,x_i)\phi(x_i)]_{a_i}
  \right\rangle\,.
\end{equation}
We should be mindful of the fact that $ ( A\cdot\der)^\xone  \barG$ is not equal to 
\begin{equation}
    A(x)\cdot\dxone\barG_{a_1\dots a_N}=
    A(x)\cdot\left\langle
[R(x,x_1)\cfd\phi(x_1)]_{a_1}\prod_{i=2}^N[R(x,x_i)\phi(x_i)]_{a_i}
  \right\rangle\,.
\end{equation}
However, from the definition of the confluent derivative of the vector-valued matrix $A$ given by
\begin{equation}
  \label{eq:y1nablaA}
   y_1\cdot \cfd A^\mu(x) \equiv
\Delta_{y_1}\Lambda^\mu_{~\nu}(x,x_1)(R(x,x_1)A^\nu(x_1) R(x_1,x))\,,\quad x_1=x+y_1\,,
\end{equation}
we can obtain
\begin{equation}
  \label{eq:RA-AR}
  R(x,x_1)A^\mu(x_1)= (A^\mu(x)+y_1\cdot\cfd A^\mu - (y_1\cdot\ucon)^\mu_{~\nu}A^\nu)R(x,x_1)\,,
\end{equation}
which allows us to commute $A$ and $R$ and relate $A(x)\cdot\dxone\barG $ and $( A\cdot\der)^\xone  \barG $:
\begin{equation}
   \label{eq:A.nabla-x1} A\cdot\dxone\barGN = (A\cdot\der)^\xone\barGN - y_1^\lambda(\cfd_\lambda A^\mu -\ucon_{\lambda\nu}^\mu A^\nu)\dxone_\mu\barGN\,.
\end{equation}
 This equation can be thought of as a generalization of Eq.~\eqref{eq:u.nabla-x1}, which is obtained by replacement $A^{\mu~b}_{~a}\to u^\mu\delta_a^b$.
Using this relation in Eq.~\eqref{eq:x1-x-y1} we obtain:
\begin{equation}
  \label{eq:A-nablaH}
  -A\cdot\cfd  \barG_{a_1\dots
      a_N}
= N \bigg[\Big( A\cdot \dyone{} + y_1^\lambda
  \left(
   \cfd_\lambda A^\mu
  - \tors^\mu_{\lambda\nu} A^\nu
\right)\dxone_\mu 
- (A\cdot\der)^\xone  \Big)_{a_1}^{~b} \barG_{ba_2\dots a_N}
\bigg]_{\overline{1\dots N}}\,.
\end{equation}
We can now put Eqs.~\eqref{eq:nablaH-sum-nablax1H-torsion} and \eqref{eq:A-nablaH} together:
\begin{multline}
    \label{eq:u-A-nablaH}
     (u-A)\cdot\cfd  \barGN
- N \left[ A\cdot \dyone{}\barGN +
  y_1^\lambda\left(
   \cfd_\lambda A^\mu
  + \tors^\mu_{\lambda\nu}(u - A)^\nu
\right)\dxone_\mu 
\barGN\right]_{\overline{1\dots N}}
\\=N\left[ ((u-A)\cdot\der)^\xone  
\barGN
+ \frac{N-1}2\Dxt (u\cdot\der)^\xone (u\cdot\der)^\xtwo\barGN
\right]_{\overline{1\dots N}}\,.
\end{multline}

So far we have not used the explicit form of the confluent connection in Eq.~\eqref{eq:bar-omega}.
Using it now, we can write:
\begin{equation}
    y_1^\lambda \tors^\mu_{\lambda\nu}(u - A)^\nu =  y_1^\lambda (-\partial_\lambda u^\mu + \aoA_\lambda u^\mu)\,,
\end{equation}
where we took into account $y_1\cdot u = A\cdot u=0$ and introduced a (spatial) vector-valued matrix
\begin{equation}
\label{eq:ao}
    \aoA_\lambda \equiv a_\lambda + 2\omega_{\lambda\nu}A^\nu\,,
\end{equation}
where 
\begin{equation}
    \omega_{\lambda\nu} \equiv \frac12\left(\Delta^\kappa_\lambda\partial_\kappa u_\nu -
\Delta^\kappa_\nu\partial_\kappa u_\lambda\right)
\end{equation}
is the vorticity, as usually defined, with 
\begin{equation}
\Delta^\kappa_\lambda\equiv \bm e_\lambda\cdot\bm e^\kappa=\delta^\kappa_\lambda+u_\lambda u^\kappa
\end{equation}
being the spatial projector in the average Landau frame.

Substituting into Eq.~\eqref{eq:u-A-nablaH} and rearranging, we find 
\begin{multline}
    \label{eq:u-A-nablaH-ao}
     (u-A)\cdot\cfd  \barGN
- N \left[\left( A +   \DA 
  \right)\cdot\dyone{}
\barGN\right]_{\overline{1\dots N}}
\\=N\left[(1+y_1\cdot \aoA)((u-A)\cdot\der)^\xone  
\barGN
+ \frac{N-1}2\Dxt (u\cdot\der)^\xone (u\cdot\der)^\xtwo\barGN
\right]_{\overline{1\dots N}}\,,
\end{multline}
where we introduced a vector-valued matrix
\begin{equation}
\label{eq:DeltaA}
    \DA^\mu\equiv y_1^\lambda\KK_\lambda^\mu\,,
    \quad\mbox{with}\quad
   \KK_\lambda^\mu\equiv 
   \cfd_\lambda A^\mu
   -\partial_\lambda u^\mu
  +\aoA_\lambda 
  A^\mu
\,.
\end{equation}
We neglected terms beyond our gradient truncation order, as in
\begin{math}
(u-A)\cdot\dxone\barG_{(N)} = ((u-A)\cdot\der)^\xone\barGN + \mathcal O(ky_1)\,,
\end{math}
which is obtained by combining  Eqs.~\eqref{eq:u.nabla-x1} and \eqref{eq:A.nabla-x1}.
Similarly, we used Eq.~\eqref{eq:x1-y1} to replace $\dxone$ with $\dyone{}$ on the LHS using the fact that $u\cdot A=0$ and $u_\mu\cfd A^\mu=0$. 

The expression for $\aoA$ in Eq.~\eqref{eq:ao} is similar to that for the inertial forces (per unit of the particle's energy) acting on a particle moving with local velocity $(-A)$ in the fluid rest frame due to acceleration and rotation (Coriolis force) of the fluid. We shall see in Section~\ref{sec:phonon} that, physically, this particle is a phonon moving in accelerating and rotating fluid.
Furthermore, as we shall see, in this physical picture the first and the second term in Eq.~\eqref{eq:DeltaA}, respectively, correspond to the force due to the spatial gradient of the phonon energy and the ``Hubble forces'' responsible, for example, for the ``red-shift'' of the phonons in an expanding medium~\cite{An:2019rhf}.

Eq.~\eqref{eq:u-A-nablaH-ao}
is the master equation we shall be using to derive equations of motion for the correlators $\barGN$. Note that we {\em have not used equations of motion \eqref{eq:u.phi-ABCD} yet}. Eq.~\eqref{eq:u-A-nablaH} is a relation between different (covariant) derivatives, $\cfd$, $\dxone$, and $\dyone{}$ we defined. As a result, Eq.~\eqref{eq:u-A-nablaH-ao} contains only $N$-point correlators. Correlators of different orders will mix in when we apply nonlinear equations of motion Eq.~\eqref{eq:u.phi-ABCD} to express the two terms in the RHS of Eq.~\eqref{eq:u-A-nablaH-ao}.

\subsection{Two-point correlators}
\label{sec:2pt}

We shall be treating fluctuations as small, since they should be small in the regime of applicability of hydrodynamics. Correspondingly, we shall use the power counting $\barGN\sim\epsilon^{N-1}$, where $\epsilon$ is the small parameter controlling the smallness of fluctuations (see Refs.~\cite{An_2021,An_2023}). 

For $N=2$, to leading order in the small parameter $\epsilon$, we only need the {\em linear} terms in the stochastic equation \eqref{eq:u.phi-ABCD}, i.e., $A$, $B$, $D$ and $C$ terms.
Specifically, for the first term on the RHS of Eq.~\eqref{eq:u-A-nablaH-ao}
we find
\begin{equation}
    (1+y_1\cdot \aoA)((u-A)\cdot\der)^\xone  
\barG = \left(B + D^{\mu\nu}\dyone{\mu}\dyone{\nu}\right)\barG\,,
\end{equation}
where $\barG\equiv \barG_{(2)}$, the $y_1\cdot \aoA$ term is negligible because $\aoA=\mathcal O(k)$, and we again used Eq.~\eqref{eq:x1-y1}.

For the last term in Eq.~\eqref{eq:u-A-nablaH-ao} we need to carefully discretize the noise correlator in Eq.~\eqref{eq:eta-eta} in the time direction:
\begin{equation}\label{eq:eta-eta-Dxt}
    \left\langle
    \eta^{\widehat c}(x_1)\eta^{\widehat d}(x_2)\right\rangle
    =2\delta^{\widehat c\widehat d}\delta^{(3)}(\bm y_1-\bm y_2)\Dxt^{-1}\quad\mbox{for $u(x)\cdot(x_1-x_2)=0$ ;}\quad \mbox{or $0$ otherwise.} 
\end{equation}
Then we find, again using Eq.~\eqref{eq:x1-y1},
\begin{equation}
    \Dxt (u\cdot\der)^\xone (u\cdot\der)^{\xtwo}\barG
    = 2 Q^{\mu\nu}\dyone{\mu}\dytwo{\nu}\delta^{(3)}(\bm y_1-\bm y_2)\,,
\end{equation}
where we introduced
\begin{equation}
    Q^{\mu\nu}_{a_1a_2}\equiv C^{\mu}_{\widehat c a_1}C^{\nu}_{\widehat d a_2}\delta^{\widehat c\widehat d}\,.
\label{eq:Q}
\end{equation}
Putting this all into the master equation \eqref{eq:u-A-nablaH-ao}
we find, with $\DA^\mu$ given by Eq.~\eqref{eq:DeltaA}, that
\begin{multline}
    \label{eq:u-A-nablaH-2pt}
     (u-A)\cdot\cfd  \barG
- 2 \left[\left( A +  \DA 
\right)\cdot\dyone{}
\barG\right]_{\overline{12}}
=
2\left[ \left(B + D^{\mu\nu}\dyone{\mu}\dyone{\nu}\right)
\barG + Q^{\mu\nu}\dyone{\mu}\dytwo{\nu}\delta^{(3)}(\bm y_1-\bm y_2)
\right]_{\overline{12}}\,,
\end{multline}
or in a more modular and compact notation:
\begin{subequations}
\label{eq:H-2pt}
\begin{equation}
   \label{eq:u.dG2} u\cdot\cfd\barG_{a_1a_2}^{(x_1,x_2)}=2\left[ \VABD_{a_1}^{~b}\barG_{ba_2}^{(x_1,x_2)}+\VQ_{a_1a_2}\delta^{(x_1,x_2)}\right]_{\overline{12}}\,.
\end{equation}
We lifted space-time variables into the superscript positions, as in Eq.~\eqref{eq:G-sup}, and, similarly, defined
\begin{equation}
\delta^{(x_1,x_2)}\equiv\delta^{(3)}(\bm y_1-\bm y_2)\,.
\end{equation}
We also defined
\begin{equation}\label{eq:V-ABD}
    \VABD \equiv \Ah + B + \Dh\,;
\end{equation}
\begin{equation}
    \Ah \equiv (A^\mu+y_1^\lambda \KK_\lambda^\mu)\dyone{\mu} + \frac1N A\cdot\cfd;
    \qquad \Dh \equiv D^{\mu\nu}\dyone{\mu}\dyone{\nu}\,,
\end{equation}
as well as
\begin{equation}
 \VQ_{a_1a_2}\equiv Q^{\mu\nu}_{a_1a_2}\dyone{\mu}\dytwo{\nu}\,.
  \label{eq:VQ}
  \end{equation}
\end{subequations}

\subsection{Three-point correlators}
\label{sec:3pt}

For $N=3$ correlators the contribution of the linear terms ($ABCD$) in Eq.~\eqref{eq:u.phi-ABCD} is similar to their contribution to $N=2$ equation \eqref{eq:u-A-nablaH-2pt}, except the $C$ terms do not contribute on account of $\langle\phi\rangle=0$. Thus,
\begin{multline}
    \label{eq:u-A-nablaH-3pt-linear}
     (u-A)\cdot\cfd  \barG_{(3)}
- 3 \left[\left( A +   
  \DA\right)\cdot\dyone{}
\barG_{(3)}\right]_{\overline{123}}
=3\left[ \left(B + D^{\mu\nu}\dyone{\mu}\dyone{\nu}\right)
\barG_{(3)}
\right]_{\overline{123}} \\ + \mbox{($A_{(3)}$, $B_{(3)}$, $D_{(3)}$, $\Dt_{(3)}$, $C_{(3)}$, $\Ct_{(3)}$ terms)} 
\,.
\end{multline}

The contribution of $A_{(3)}$ requires most work, so we shall consider it separately. Using Eq.~\eqref{eq:u.phi-ABCD} we can write for the relevant part of the first term on the RHS of Eq.~\eqref{eq:u-A-nablaH-ao}:
\begin{equation}
 \label{eq:u-A-A3}   ((u-A)\cdot\der)^\xone  
\barG_{a_1a_2a_3} = \left\langle R_{a_1}^{~a}(x,x_1)A^{\mu~bc}_{~a}(x_1)\subtr{\cfd_\mu\phi_b(x_1)\phi_c(x_1)}\prod_{i=2}^3[R\phi(x_i)]_{a_i}\right\rangle
+\dots
\end{equation}
where in the correlator we use the shorthand notation introduced in \eqref{eq:H}. The ellipsis denotes contribution of terms without $A_{(3)}$ which we shall consider later. We need to reexpress the $A_{(3)}$ term in Eq.~\eqref{eq:u-A-A3} in terms of the confluent correlator and its derivative multiplied by $A_{(3)}(x)$. We shall use the same trick as in Eqs.~\eqref{eq:y1nablaA} and \eqref{eq:RA-AR} for commuting $A$ and $R$ matrices, applying it now to $A_{(3)}$. By definition
\begin{equation}
    y_1\cdot\cfd A_{~a_1}^{\mu~bc}(x)
    = \Delta_{y_1} 
    \Lambda^\mu_{~\nu}(x,x_1)R_{a_1}^{~a'}(x,x_1)A_{~a'}^{\nu~b'c'}(x_1)
    R_{b'}^{~b}(x_1,x) R_{c'}^{~c}(x_1,x)
\end{equation}
(as before, $x_1=x+y_1$), thus,
\begin{multline}\label{eq:RARR}
    R_{a_1}^{~a}(x,x_1)A_{~a}^{\mu~bc}(x_1)
    = \Lambda^\mu_{~\nu}(x_1,x) ( A_{~a_1}^{\nu~b'c'}(x) + y_1\cdot\cfd A_{~a_1}^{\mu~b'c'}) 
    R_{b'}^{~b}(x,x_1) R_{c'}^{~c}(x,x_1)
    \\=
\left[ \left(A^{\mu}_{(3)}(x) + y_1\cdot\cfd A^{\mu}_{(3)} - (y_1\cdot\ucon)^{\mu}_{~\nu} A^{\nu}_{(3)}\right) 
    R(x,x_1) R(x,x_1)\right]_{a_1}^{~bc}\,.
\end{multline}
In the last expression we suppressed the indices where this does not create an ambiguity. Substituting into Eq.~\eqref{eq:u-A-A3} we find for the $A_{(3)}$ terms
\begin{multline}
 \label{eq:u-A-A3RR}   ((u-A)\cdot\der)^\xone  
\barG_{a_1a_2a_3} =   \left[A^{\mu}_{(3)} + y_1\cdot\cfd A^{\mu}_{(3)} - (y_1\cdot\ucon)^{\mu}_{~\nu} A^{\nu}_{(3)}     \right]_{a_1}^{~bc}\\\times
\left\langle\subtr{[R\cfd_\mu\phi(x_1)]_b[R\phi(x_1)]_c}\prod_{i=2}^3[R\phi(x_i)]_{a_i}\right\rangle + \dots\,.
\end{multline}

\begin{figure}[h]
\includegraphics[height=8em]{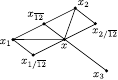}

\caption[]{Geometric representation of the fact that the midpoint of $x_1$ and $x_2$ is not $x$, but $x_\av{12}$. The correlator in the RHS of Eq.~\eqref{eq:Hx-H} is defined between points $x_{1/\av{12}}$ and $x_{2/\av{12}}$ whose midpoint {\em is\/} $x$.}
\label{fig:x-x12}
\end{figure}

We can now use Wick's theorem to reexpress the correlator in terms of products of two correlators obtained by contracting each of the fields inside $\subtr{\dots}$ with each subset of the fields outside of it:\footnote{In the power counting we use, $\barGN\sim\epsilon^{N-1}$, guaranteed by the smallness of fluctuations, $2\barG_{12}\barG_{13}$ is the leading term in $\langle\subtr{\phi_1\phi_1}\phi_2\phi_3\rangle$, and is of the same order as $\barG_{(3)}\sim\epsilon^2$.}
\begin{multline}
  \label{eq:G3-GG} \left\langle\subtr{[R\cfd_\mu\phi(x_1)]_b[R\phi(x_1)]_c}\prod_{i=2}^3[R\phi(x_i)]_{a_i}\right\rangle \\=  \left\langle[R(x,x_1)\cfd_\mu\phi(x_1)]_b[R(x,x_2)\phi(x_2)]_{a_2}\right\rangle
   \Big\langle[R(x,x_1)\phi(x_1)]_c[R(x,x_3)\phi(x_3)]_{a_3}\Big\rangle + (2\leftrightarrow3)
\\ \equiv \dxone\barG^{(x_1,x_2)}_{ba_2(x)}\barG^{(x_1,x_3)}_{ca_3(x)}
   \,.
\end{multline}
In the last line we introduced a generalization of the confluent correlator, such that the point $x$, with respect to which the correlator is covariant and equal-time, is not  the same as the midpoint. We shall refer to such a correlator, which obeys the same relations derived in Sections \ref{sec:cc-correlator} and \ref{sec:cc-derivs}, as {\em dislocated\/} confluent correlator and distinguish it by the additional subscript $(x)$. In particular, the midpoint of a three-point correlator, $x=(x_1+x_2+ x_3)/3$, as defined in Eq.~\eqref{eq:x}, is not the midpoint of $x_1$ and $x_2$. We denote the midpoint of $x_1$ and $x_2$ by $x_\av{12}\equiv (x_1+x_2)/2$ as shown in Fig.\ref{fig:x-x12}. The midpoint $x_\av{12}$ is separated from $x$ by the dislocation vector
\begin{equation}
    y_\av{12}\equiv x_\av{12}-x. 
\end{equation}
On the other hand, the midpoint of 
\begin{equation}
    x_{1/\av{12}}\equiv x_1-y_\av{12}\equiv x+ y_{1/\av{12}}\quad\mbox{and}\quad x_{2/\av{12}}\equiv x_2-y_\av{12}\equiv x + y_{2/\av{12}}
\label{eq:x1/12}
\end{equation}
 {\em is} $x$ (see Fig.\ref{fig:x-x12}).

Substituting Eq.~\eqref{eq:G3-GG} into Eq.~\eqref{eq:u-A-A3RR} we shall consider the terms arising from $A_{(3)}^\mu$, $y_1\cdot\cfd A_{(3)}^\mu$, and $y_1\cdot\ucon^\mu_{~\nu} A_{(3)}^\nu$ in turn. 
 Index $\mu$ of $A^\mu_{(3)}(x)$ is spatial in the frame $u(x)$, i.e., $A_{(3)}\cdot u=0$.
Therefore, we can use Eq.~\eqref{eq:x1-x-y1}:
\begin{multline}
\label{eq:A3-dxone}
    A^{~bc}_{a_1}
    \cdot\dxone\barG_{ba_2(x)}^{(x_1,x_2)}
    =  A_{a_1}^{~bc}\cdot\left(\dyone{} 
+   \frac12\cfd\right) \barG_{ba_2(x)}^{(x_1,x_2)}
    +A^{\nu~b_1c}_{~a_1}\left[\ucon_{\nu\lambda}^\mu y_{1}^\lambda
     \dxone_\mu\barG_{b_1b_2(x)}^{(x_1,x_2)}\right]_{\overline{12}}\delta^{b_2}_{a_2(x)}
     \\=A^{\nu~bc}_{~a_1}\left(\dyone{\nu} 
+   \frac12\cfd_\nu + \ucon_{\nu\lambda}^\mu y_{1/\av{12}}^\lambda
     \dxone_\mu\right)\barG_{ba_2(x)}^{(x_1,x_2)}
\,,
\end{multline}
where in the last line we used $y_1+y_2=y_\av{12}$ and $\dxone+\dxtwo=\mathcal O(k)$ (which is negligible here) for the $N=2$ correlator. Vector $y_{1/\av{12}}= y_1-y_\av{12}$ is defined in Eq.~\eqref{eq:x1/12}.

For the second term in Eq.~\eqref{eq:u-A-A3RR}, we use Eq.~\eqref{eq:x1-y1},
\begin{equation}\label{eq:ydA}
    \left(y_1\cdot\cfd A^{\mu~bc}_{~a_1}\right)\dxone_\mu\barG_{ba_2} =
    \left(y_1\cdot\cfd A^{\mu~bc}_{~a_1}\right)\dyone{\mu}\barG_{ba_2}\,,
\end{equation}
since the index $\mu$ is spatial and only the leading term contributes within the order of our gradient truncation. 

Substituting Eqs.~\eqref{eq:G3-GG}, \eqref{eq:A3-dxone}, and \eqref{eq:ydA} into Eq.~\eqref{eq:u-A-A3RR} we find
for the relevant part of this equation
\begin{multline}\label{eq:A3-y1-dH}
\left[A^{\mu}_{(3)} + y_1\cdot\cfd A^{\mu}_{(3)} - (y_1\cdot\ucon)^{\mu}_{~\nu} A^{\nu}_{(3)}     \right]_{a_1}^{~bc}\dxone_\mu\barG_{ba_2(x)}^{(x_1,x_2)}
\\= \left[A^{\mu}_{(3)}\left(\dyone{\mu} 
+   \frac12\cfd_\mu + \left(\tors_{\mu\lambda}^\nu y_{1}^\lambda-\ucon_{\mu\lambda}^\nu y_\av{12}^\lambda\right)
     \dxone_\nu\right)+(y_1\cdot\cfd) A_{(3)}\cdot\dyone{}\right]^{~bc}_{a_1}\barG_{ba_2(x)}^{(x_1,x_2)}\,.
\end{multline}
We can express dislocated correlator in terms of the ordinary confluent correlator (truncating terms beyond our gradient expansion precision):
\begin{equation}
   \label{eq:Hx-H} H^{(x_1,x_2)}_{a_1a_2(x)} = \left(1+y_\av{12}\cdot\left(\dxone+\dxtwo\right)\right)H^{(x_{1/\av{12}},x_{2/\av{12}})}_{a_1a_2}\,,
\end{equation}
where we used the definition of confluent derivative to express 
\begin{equation}
    R(x,x_1)\phi(x_1)=R(x,x_{1/\av{12}})\left(1+y_\av{12}\cdot\cfd\right)\phi(x_{1/\av{12}})\,.
\end{equation}
Similarly,
\begin{equation}
    \dxone_\mu H^{(x_1,x_2)}_{a_1a_2(x)} = \left(\dxone_\mu+y_\av{12}\cdot\left(\dxone+\dxtwo\right)\dxone_\mu + y_\av{12}^\lambda\ucon^\nu_{\lambda\mu}\dxone_\nu \right)H^{(x_{1/\av{12}},x_{2/\av{12}})}_{a_1a_2}
\end{equation}
where we again used the definition of the confluent derivative (this time, of a Lorentz (co-)vector) to express 
\begin{multline}
R(x,x_1)\cfd\phi(x_1)=\Lambda(x_1,x_{1/\av{12}})R(x,x_{1/\av{12}})\left(1+y_\av{12}\cdot\cfd\right)\cfd\phi(x_{1/\av{12}})\\=R(x,x_{1/\av{12}})\left(1+y_\av{12}\cdot(\cfd+\ucon)\right)\cfd\phi(x_{1/\av{12}})\,.
\end{multline}
Substituting into Eq.~\eqref{eq:A3-y1-dH} --- only the first, leading term is sensitive to the dislocation within the gradient expansion order we are truncating at --- we find
\begin{multline}\label{eq:A3-y1-dH-2}
\left[A^{\mu}_{(3)} + y_1\cdot\cfd A^{\mu}_{(3)} - (y_1\cdot\ucon)^{\mu}_{~\nu} A^{\nu}_{(3)}     \right]_{a_1}^{~bc}\dxone_\mu\barG_{ba_2(x)}^{(x_1,x_2)}
\\= \left[A^{\mu}_{(3)}\left(\dyone{\mu} + y_\av{12}\cdot\left(\dxone+\dxtwo\right)\dyone{\mu}
+   \frac12\cfd_\mu + \tors_{\mu\lambda}^\nu y_{1/\av{12}}^\lambda
     \dxone_\nu\right)+(y_1\cdot\cfd) A_{(3)}\cdot\dyone{}\right]^{~bc}_{a_1}\barG_{ba_2}^{(x_{1/\av{12}},x_{2/\av{12}})}\,.
\end{multline}
The indices $\mu\lambda$ of the confluent torsion $\tors_{\mu\lambda}^\nu$ are contracted with spatial vectors $A^\mu_{(3)}$ and $y_1^\lambda$, therefore, using Eq.~\eqref{eq:bar-omega}, we can replace $\tors^\nu_{\mu\lambda}\to 2\omega_{\lambda\mu}u^\nu$. The resulting temporal derivative can be replaced using the leading-order term in the equation of motion for $\phi$, Eq.~\eqref{eq:u.phi-ABCD}: 
\begin{equation}
    u\cdot\cfd\phi = A\cdot\cfd\phi+\mathcal O(k) \quad\Rightarrow\quad
   u\cdot\dxone\barG = A\cdot
    \dyone{}\barG+\mathcal O (k)\,.
\label{eq:u.d=A.d}
\end{equation}

Putting all this together into Eq.~\eqref{eq:u-A-A3RR} and then into master equation \eqref{eq:u-A-nablaH} we find for the $A_{(3)}$ terms in Eq.~\eqref{eq:u-A-nablaH-3pt-linear}:
\begin{multline}
\label{eq:1Yu-A-du}
    \left[(1+y_1\cdot Y)((u-A)\cdot\der)^\xone  
\barG_{a_1a_2a_3}^{(x_1,x_2,x_3)}\right]_{\overline{123}} \\= 2\Bigg[\left(\left(\left(
A_{(3)}+\DA_{(3)}\right)\cdot\dyone{}+ \frac12A_{(3)}\cdot\cfd\right)\otimes1 + \hat\Pi_{(3)}\right)_{a_1}^{~bc} \args{\barG_{ba_2}^{(x_{1/\av{12}},x_{2/\av{12}})},\barG_{ca_3}^{(x_{1/\av{13}},x_{3/\av{13}})}}
\Bigg]_{\overline{123}}
+ \dots\,,
\end{multline}
where
\begin{equation}
    \DA_{(3)}^\mu\equiv y_1^\lambda\left(\cfd_\lambda A_{(3)}^\mu + \aoA_\lambda A_{(3)}^\mu+2\omega_{\lambda\nu}A_{(3)}^\nu A^{\mu}\right)\,. 
\end{equation}
Note, that $\DA_{(3)}$ is simply a derivative of matrix $\DA$ defined in Eq.~\eqref{eq:DeltaA}:
\begin{equation}
    \label{eq:DeltaA3}
    \DA^{\mu~bc}_{~a}=\DA^{\mu~b,c}_{~a}\,,
\end{equation}
where we used Eq.~\eqref{eq:A3A'}.

Note that all correlators in Eq.~\eqref{eq:1Yu-A-du}
have the same midpoint --- point $x$. This would not be the case if the arguments of the  correlators were $(x_1,x_2)$ and $(x_1,x_3)$. However, the difference only matters in the leading term --- the first term, $A_{(3)}\cdot\dyone{}$. The operator denoted by $\hat\Pi_{(3)}$
represents the effect of the dislocation between the midpoints $x_\av{12}$ and $x_\av{13}$ of the 2-point correlators and the midpoint $x=x_\av{123}$ of the three-point correlator:
\begin{multline}
    \hat\Pi_{(3)}
    \equiv A^{\mu}_{(3)}\Big[y_\av{12}^\lambda\left(\left(\dxone_\lambda+\dxtwo_\lambda\right)\dxone_\mu+\tors_{\lambda\mu}^\nu\dxone_\nu\right)\otimes1 + \dxone_\mu\otimes y_\av{13}\cdot\left(\dxone+\dxtwo\right)\Big] 
\\ = A^{\mu}_{(3)}\Big[y_\av{12}^\lambda\left(\left(\cfd_\lambda+2\ucon_{\lambda\kappa}^\nu y_{1/\av{12}}^\kappa\dyone{\nu}\right)
\dyone{\mu}+2\omega_{\mu\lambda}A\cdot\dyone{}\right)\otimes1 +  \dyone{\mu}\otimes y_\av{13}^\lambda\left(\cfd_\lambda+2\ucon_{\lambda\kappa}^\nu y_{1/\av{13}}^\kappa\dyone{\nu}\right)\Big]
\end{multline}
where in the last line we expressed all derivatives in terms of the confluent derivative $\cfd$ with respect to the midpoint and the balanced derivatives $\dyone{}$ using Eqs.~\eqref{eq:x1-x-y1} and \eqref{eq:u.d=A.d}. This facilitates Wigner transform we discuss in the next section.

The ellipsis in Eq.~\eqref{eq:1Yu-A-du} denotes terms without $A_{(3)}$, i.e., $B$, $D$, $B_{(3)}$, 
$D_{(3)}$, $\Dt_{(3)}$ terms. These terms are easier to work out because they are already at the highest order in gradient expansion and all covariant connection matrices $R$ and $\Lambda$ can be replaced by a unit matrix without loss of precision. The difference between, e.g., $B(x_1)$ and $B(x)$ is similarly negligible. For the 
$B_{(3)}$, 
$D_{(3)}$, $\Dt_{(3)}$ we thus find
\begin{multline}
    \left[(1+y_1\cdot Y)((u-A)\cdot\der)^\xone  
\barG_{a_1a_2a_3}^{(x_1,x_2,x_3)}\right]_{\overline{123}} \\= 2\left[\left( B_{(3)} + D_{(3)}^{\mu\nu}\dyone{\mu}\dyone{\nu}\otimes1+\Dt_{(3)}^{\nu\mu} \dyone{\mu}\otimes\dyone{\nu} \right)_{a_1}^{~bc} \args{\barG_{ba_2}^{(x_1,x_2)},\barG_{ca_3}^{(x_1,x_3)}}\right]_{\overline{123}} + \dots\,.
\end{multline}

The last term in Eq.~\eqref{eq:u-A-nablaH} gives rise to $CC$ type terms, which are similarly easier to deal with because they are already at the highest retained order in gradients. We find, using equations of motion \eqref{eq:u.phi-ABCD} and regulated noise correlator Eq.~\eqref{eq:eta-eta-Dxt}:
\begin{multline}
\Dxt\left[ (u\cdot\der)^\xone (u\cdot\der)^\xtwo\barG_{a_1a_2a_3}^{(x_1,x_2,x_3)}
\right]_{\overline{123}} \\ = 
2\left[\left(
Q_{(3)}^{\mu\nu} \otimes\dyone{\mu}\dytwo{\nu} + \tilde Q_{(3)}^{\mu\nu}\dyone{\mu}\otimes\dytwo{\nu}
\right)_{a_1a_2}^{~~b}\args{\barG_{ba_3}^{(x_1,x_3)},
\delta^{(x_1,x_2)}} \right]_{\overline{123}}
\end{multline}
where 
\begin{equation}
    Q^{\mu\nu~~b}_{a_1a_2} \equiv 2C^{\mu~,b}_{\widehat c a_1} C^\nu_{\widehat d a_2} \delta^{\widehat c_1\widehat c_2}
    \quad\mbox{and}\quad
    \Qt^{\mu\nu~~b}_{a_1a_2} \equiv 2\Ct^{\mu~b}_{\widehat c a_1} C^\nu_{\widehat d a_2}\delta^{\widehat c_1\widehat c_2}
    \,.
    \label{eq:Q3C'C}
\end{equation}

Putting everything together into Eq.~\eqref{eq:u-A-nablaH} we can write the three-point function evolution equation as follows:
\begin{subequations}
    \label{eq:H-3pt}
\begin{multline}
 u\cdot\cfd\barG_{a_1a_2a_3}^{(x_1,x_2,x_3)}
 =  3\Big[ \VABD_{a_1}^{~b}\barG_{ba_2a_3}^{(x_1,x_2,x_3)}  + 2\VABD_{a_1}^{~bc} \args{\barG_{ba_2}^{(x_{1/\av{12}},x_{2/\av{12}})}
   ,\barG_{ca_3}^{(x_{1/\av{13}},x_{3/\av{13}})} 
    } \\+ 2\VQ_{a_1a_2}^{~~~b}\args{\barG_{ba_3}^{(x_{1/\av{13}},x_{3/\av{13}})},\delta^{(x_1,x_2)}
    }\Big]_{\overline{123}} \label{eq:u.dG3}
\end{multline}
where we used notations similar to those in Eq.~\eqref{eq:u.dG2}
We also defined bilinear (vertex) operators and hypermatrices (with three indices) acting on two correlators:
\begin{equation}
    \VABD_{(3)}\equiv \Ah_{(3)} + B_{(3)} + \Dh_{(3)} + \hat\Pi_{(3)}\,,
\label{eq:VABD3}
\end{equation}
where, using Eqs.~\eqref{eq:A3A'} and \eqref{eq:BDC'},
\begin{equation}
    \Ah_a^{~bc} \equiv \left(\left(A^\mu+y_1^\lambda \KK_\lambda^\mu\right)\dyone{\mu}+\frac12A\cdot\cfd\right)_a^{~b,c}\otimes1,
\end{equation}
\begin{equation}
    \Dh_a^{~bc} \equiv D_{~~a}^{\mu\nu~b,c}\dyone{\mu}\dyone{\nu}\otimes1 + \Dt_{~~a}^{\nu\mu~bc}\dyone{\mu}\otimes\dyone{\nu}\,,
\end{equation}
and
\begin{equation}
    \VQ_{a_1a_2}^{~~~b}\equiv Q^{\mu\nu~b}_{a_1a_2}\otimes \dyone{\mu}\dytwo{\nu} + \Qt_{a_1a_2}^{\mu\nu~b}\,\dyone{\mu}\otimes\dytwo{\nu}\,.
\end{equation}
\end{subequations}

\section{Wigner transform}
\label{sec:wigner}

To take full advantage of the separation of scales given by Eq.~\eqref{eq:kq-hierarchy} it is convenient to represent the hydrodynamic correlators $\barGN$ using their Fourier-Wigner transform, $\WN$, with respect to the spatial coordinates $\bm y_i$. 
Such a generalized Wigner transform was introduced in Ref.~\cite{An:2019rhf}:
\begin{equation}
    \WN\left(x;\{\bm q_i\}_{i=1}^N\right) \equiv \int \barGN\left(x;\{\bm y_i\}_{i=1}^N\right)\delta^{(3)}\left(\frac1N\sum_{i=1}^N\bm y_i\right)\prod_{i=1}^N e^{-i\bm q_i\cdot\bm y_i}d^3\bm y_i\equiv \WT_{(N)}\left[\barG_{(N)}\right] \,.
\label{eq:WG}
\end{equation}
Similarly to $\barGN$, Wigner function $\WN$ is SO(3) covariant in the sense of Eqs.~\eqref{eq:H-TH},~\eqref{eq:H-TH-y}, i.e.,
\begin{equation}
    W_{a_1,\dots,a_N}\left(x;\{\bm q_i\}_{i=1}^N\right) \to \left(\prod_{i=1}^{N} T_{a_i}^{~b_i}(x)\right)W_{b_1\dots b_N}\left(x;\{T(x)^{-1}\bm q_i\}_{i=1}^N\right)\,.
    \label{eq:W-TW}
\end{equation}
The SO(3) covariant  confluent derivative of $W$ is the Wigner transform of the confluent derivative $\cfd\barG$ and is given by (see Eq.~\eqref{eq:W-TW} and also Ref.~\cite{An_2023})
\begin{equation}
    \cfd_\mu W_{a_1\dots a_N} = \partial_\mu W_{a_1\dots a_N} + N\left[\mathring\omega_{\mu b}^a q_{1a}\frac{\partial}{\partial q_{1b}}W_{a_1\dots a_N} - \mathring\omega_{\mu a_1}^{b}W_{ba_2\dots a_N}  \right]_\oN = \WT_{(N)}[\cfd\barGN]\,.
\end{equation}
The Wigner transforms of $\bm y_i$ and of the balanced derivative $\bdyi$ are simple, due to Eq.~\eqref{eq:d-sum-y}:
\begin{equation}
   \WT_{(N)} \left[\bm y_1 \barGN\right] = i\frac{\partial}{\partial\bm q_1}\WN\,;\quad
   \WT_{(N)} \left[\bdyone \barGN\right]=
   i\bm q_1'\WN\,, \quad\mbox{where}\quad \bm q_1'\equiv \bm q_1-\frac1N\sum_{i=1}^N\bm q_i\,.
\label{eq:WTyG}
\end{equation}
It is easy to see from Eq.~\eqref{eq:WG} that $\WN$ is invariant with respect to the shift of all $\bm q_i$ by the same vector. In particular,
\begin{equation}
\WN\left(x;\{\bm q_i'\}_{i=1}^N\right) = \WN\left(x;\{\bm q_i\}_{i=1}^N\right).
\end{equation}
Therefore, it is sufficient to consider only sets of $\bm q_i$ which satisfy $\sum_{i=1}^N\bm q_i=0$ and, therefore, $\bm q_i'=\bm q_i$. 

Similarly to the Fourier transform, the Wigner transform of the delta-function is unity, i.e., 
\begin{equation}
    \WT_{(2)}\left[\delta^{(x_1,x_2)}\right]=1\,.
\end{equation}
Another relation, especially useful for three-point (non-Gaussian) correlator equation, is the $N=3$ Wigner transform of a product of two  (undislocated) correlators sharing the midpoint $x$. 
\begin{equation}
    \WT_{(3)}\left[ \barG^{(x_{1/\av{12}},x_{2/\av{12}})}_{ba_2}\barG^{(x_{1/\av{13}},x_{3/\av{13}})}_{ca_3}\right]=W^{(-q_2',q_2')}_{ba_2}(x)W^{(-q_3',q_3')}_{ca_3}(x)\,.
\end{equation}
Note that the primed variables $q_i'$ appearing in this Wigner transform are trivially independent of the shift of {\em all three\/} $q_i'$ by the same vector.

The easiest way to derive relations like these is to 
express the confluent correlators in terms of Wigner functions using the inverse Wigner transform,
\begin{equation}
    \barG^{(x_1,\dots,x_N)}_{(N)} = \int W^{(\bm q_1,\dots,\bm q_N)}_{(N)} \delta^{(3)}\left(\sum_{i=1}^N\bm q_i\right)\prod_{i=1}^N e^{i\bm q_i\bm y_i}\frac{d^3\bm q_i}{(2\pi)^3}\,,
\end{equation}
use the integral representation of the $\delta$-function appearing in the definition of the Wigner transform,
\begin{equation}
    \delta^{(3)}\left(\frac1N\sum_{i=1}^N\bm y_i\right)=
    \int_{\bm q}\exp\left\{\frac iN\sum_{i=1}^N\bm q\cdot\bm y_i\right\}\,,
\end{equation}
integrate over $\bm y$'s first and use the resulting $\delta$ functions to eliminate the integrals over the wave-vectors.

\subsection{Two-point Wigner function equation}

Applying Wigner transform in Eq.~\eqref{eq:WG} to the two-point correlator equation \eqref{eq:u.dG2} we thus find
\begin{subequations}
\begin{equation}
    u\cdot\cfd W_{a_1a_2}=2\left[ \VABDW_{a_1}^{~b} W_{ba_2} + \VQW_{a_1a_2}
\right]_{\overline{12}}\,,
\label{eq:u.dW2}\end{equation}
where the action of the operator $\VABDW$ is again given by Eq.~\eqref{eq:V-ABD} with operators $\Ah$ and $\Dh$ acting on Wigner functions as
\begin{equation}\label{eq:Ah}
        \Ah W^{(q_1,q_2)} \equiv \left(A\cdot\left(iq_1+ \frac1N\cfd\right)-\frac{\partial}{\partial q_{1\lambda}} \KK_\lambda^\mu q_{1\mu}' \right)W^{(q_1,q_2)};
   \end{equation}
\begin{equation}
    \quad \Dh W^{(q_1,q_2)} \equiv - D(q_1,q_1)W^{(q_1,q_2)} \,,\quad\mbox{and}\quad
    \VQW 
    \equiv -Q(q_1,q_2) \,.
\label{eq:VQqq}
\end{equation}
\label{eq:W2-eqs}
To reduce the index clutter we suppressed Lorentz indices by using notation for bilinear operators or quadratic forms in the definitions of $\Dh$ and $\VQ$ above.
 In this notation
\begin{equation}
    D(q_1,q_1)\equiv D^{\mu\nu}q_{1\mu}q_{1\nu}\,,\quad Q(q_1,q_2)\equiv Q^{\mu\nu}q_{1\mu}q_{2\nu}\,.
\label{eq:Dqq}\end{equation}
\end{subequations}

\subsection{Three-point Wigner function equation}

Applying Wigner transform to Eq.~\eqref{eq:u.dG3} we find
\begin{subequations}
\begin{equation}
    u\cdot\cfd W_{a_1a_2a_3}^{(q_1,q_2,q_3)}
 =  3\left[ \VABDW_{a_1}^{~b}W_{ba_2a_3}^{(q_1,q_2,q_3)}  + 2\VABDW_{a_1}^{~bc} \args{W_{ba_2}^{(-q_2',q_2')}
   ,W_{ca_3}^{(-q_3',q_3')}}
     + 2\VQW_{a_1a_2}^{~~~b}W_{ba_3}^{(-q_3',q_3')}\right]_{\overline{123}}\,.
\label{eq:u.dW3}
\end{equation}
The bilinear operator $\VABDW_{(3)}$ is the result of Wigner transform of the operator in Eq.~\eqref{eq:VABD3},
where the action of $\Ah_{(3)}$ 
on the pair of the Wigner functions is given by:
\begin{multline}\label{eq:Ah3W}
      \Ah_{a_1}^{~bc} \args{W_{ba_2}^{(-q_2',q_2')},W_{ca_3}^{(-q_3',q_3')}} \equiv \left(A\cdot\left(i(-q_2')+ \frac12\cfd\otimes1\right)-\frac{\partial}{\partial q_{1\lambda}} \KK_\lambda^\mu \left(-q_{2\mu}'\right) \right)_{a_1}^{~b,c}\args{W_{ba_2}^{(-q_2',q_2')},W_{ca_3}^{(-q_3',q_3')}},
\end{multline}
where the $q_i'$ can be set to $q_i$ {\em only after\/} the derivative with respect to $q_1$ has been taken, since both $q_2'$ and~$q_3'$ depend on $q_1$: $\partial q_{2\mu}'/\partial q_{1\lambda}=\partial q_{3\mu}'/\partial q_{1\lambda}=-(1/3)\delta^\lambda_\mu$, according to Eq.~\eqref{eq:WTyG}.
The operator $\Dh_{(3)}$ in $\VABDW_{(3)}$ is given by
\begin{equation}
    \Dh_{a_1}^{~bc} \args{W_{ba_2}^{(-q_2',q_2')},W_{ca_3}^{(-q_3',q_3')}} \equiv - \left(D_{a_1}^{~b,c}(q_2',q_2') + \Dt_{a_1}^{~bc}(q_3',q_2')\right) \args{W_{ba_2}^{(-q_2',q_2')},W_{ca_3}^{(-q_3',q_3')}} \,,
\end{equation}
while
\begin{equation}
    \VQW_{a_1a_2}^{~~~b}W_{ba_3}^{(-q_3',q_3')}\equiv\left(
    Q_{a_1a_2}^{~~~b}(q_2',q_2')+\Qt_{a_1a_2}^{~~~b}(q_3',q_2')
    \right)
    W_{ba_3}^{(-q_3',q_3')}\,.
\label{eq:VQ3W}\end{equation}
The dislocation term is given by
\begin{multline}
    \hat\Pi_{(3)}\args{W^{(-q_2',q_2')},W^{(-q_3',q_3')}} \\= A^{\mu}_{(3)}\Bigg[\frac12\frac{\partial}{\partial q_{3\lambda}}\left(\left(\cfd_\lambda\otimes1-\ucon_{\lambda\kappa}^\nu\left(\frac\partial{\partial q_1}-\frac\partial{\partial q_2}\right)^\kappa (-q_{2\nu}')\right)(-q_{2\mu}')
+2\omega_{\mu\lambda}A\cdot(-q_{2}')\right) \\ + \frac12\frac\partial{\partial q_{2\lambda}} \left(1\otimes\cfd_\lambda-\ucon_{\lambda\kappa}^\nu\left(\frac\partial{\partial q_1}-\frac\partial{\partial q_3}\right)^\kappa(-q_{3\nu}')\right)(-q_{2\mu}')\Bigg]\args{W^{(-q_2',q_2')},W^{(-q_3',q_3')}} 
\end{multline}
Note that $(\partial/\partial q_1-\partial/\partial q_3)q_2'=0$.
\label{eq:W3-eqs}
\end{subequations}

\begin{figure}[h]
    \centering
    \includegraphics[width=0.8\linewidth]{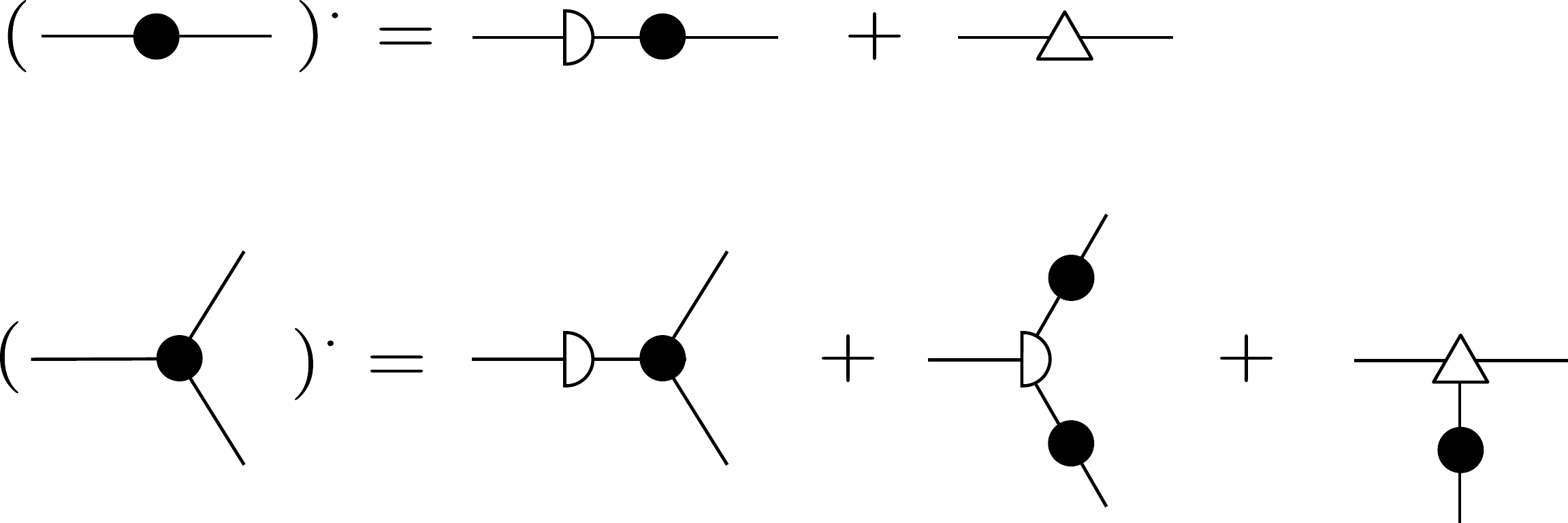}
    \caption{Diagrammatic representation of Eqs.~\eqref{eq:u.dW2} and \eqref{eq:u.dW3}.}
    \label{fig:gamma-diagrams}
\end{figure}

Equations \eqref{eq:u.dW2} and \eqref{eq:u.dW3} have intuitive diagrammatic representation as shown in Fig.\ref{fig:gamma-diagrams}. The topology of the diagrams is universal, i.e., it is the same as in the simpler problem of non-Gaussian fluctuations in a system with a discrete set of stochastic variables introduced in Ref.~\cite{An_2021}, as well as the problem of stochastic nonlinear diffusion studied in that paper. The novelty of our results here is in the determination of the vertex operators for the {\em full} set of hydrodynamic variables, including {\em non-Gaussian} local momentum fluctuations.

\section{Equilibrium conditions}
\label{sec:equilibrium}

The essential property of dissipative hydrodynamics is the relaxation of the fluid it describes to equilibrium. We would like to make sure that this property is reflected in the equations for the correlators we derived from the stochastic equation \eqref{eq:u.phi-ABCD}. As we shall see, the existence of an equilibrium solution requires that the coefficients $ABCD$ obey certain conditions. We shall determine these conditions in this Section. We shall then check, in the next Section, that these conditions are fulfilled if the equation  \eqref{eq:u.phi-ABCD} is derived from {\em hydrodynamics}, i.e., a theory which obeys conservation laws as well as the second law of thermodynamics.

We shall consider a homogeneous equilibrium state. In such a state all spatial or temporal derivatives, such as $\cfd$, vanish. In addition to this global equilibration, the fluid's local equilibration means that the correlators approach their equilibrium values, given by thermodynamics. The equilibrium correlators are local (point-like on the relevant hydrodynamic scales) and corresponding Wigner functions, which we denote by $\Weq_{(N)}$, are $\bm q$ independent. We would like to make sure that these equilibrium values satisfy our equations.

We start with Eq.~\eqref{eq:u.dW2} for
$N=2$. In equilibrium, setting all gradients to zero, 
we find 
\begin{equation}\label{eq:ADWQ}
0=\left[\big(iq_1\cdot A-D \args{q_{1}, q_{1}}\big)_{a_1}^{~b_1}\Weq_{b_1a_2} - Q_{a_1a_2}\args{q_{1}, q_{2}}\right]_{\overline{12}}\,,
\end{equation}
where to reduce index clutter, we suppressed Lorentz indices by using notation for bilinear operators in Eq.~\eqref{eq:Dqq}. 
The coefficient of each linearly independent $q_i$ must vanish independently. Since, for $N=2$ there is only one linearly independent vector $q_i$, due to $q_1+q_2=0$, let us choose it to be $q_1$ and express $q_2$ as $-q_1$. Then, vanishing of the terms linear in $q_1$ requires:
\begin{equation}\label{eq:AW-12}
    \bm A_{a_1}^{~b}\Weq_{ba_2}-(1\leftrightarrow2)=0\,,\quad \mbox{where}\quad \bm A\equiv\bm e_\mu A^\mu\,.
\end{equation}
In other words, the vector-valued matrix
\begin{equation}
\label{eq:Omega-def}
    \bm \Omega_{a_1a_2}\equiv \bm A_{a_1}^{~b}\Weq_{ba_2}
   \,, \quad\mbox{or}\quad\bm \Omega=\bm A\Weq\,,
\end{equation}
must be symmetric: 
\begin{equation}\label{eq:Omega12}
    \bm\Omega_{a_1a_2}=\bm\Omega_{a_2a_1}\,,\quad\mbox{or}\quad
    \bm\Omega=\bm\Omega^T.
\end{equation}

There is only one linearly independent bilinear among $\args{q_i, q_j}$, $i=1,2$ (again, because $q_1+q_2=0$). We choose that to be $\args{q_1, q_2}$ and express all other bilinears in terms of it (e.g., $\args{q_1, q_1}=-\args{q_1, q_2}$). We then require the coefficient of $q_{1\mu} q_{2\nu}$ to be zero in Eq.~\eqref{eq:ADWQ}. This leads to
\begin{equation}\label{eq:DWQ}
E^{\mu\nu}_{a_1a_2}+E^{\nu\mu}_{a_2a_1}=0\,,\quad\mbox{where}\quad
E^{\mu\nu}_{a_1a_2}\equiv 
D_{~~a_1}^{\mu\nu~b}\Weq_{ba_2}-Q^{\mu\nu}_{a_1a_2}\,.
\end{equation}
This condition is reminiscent of a familiar relationship between the diffusion coefficient ($D$), susceptibility~($\Weq$), and conductivity ($Q$).
In Section \ref{sec:hydro-conserved} we shall verify explicitly that the conditions expressed by Eqs.~\eqref{eq:AW-12} (or Eq.~\eqref{eq:Omega12}) and Eq.~\eqref{eq:DWQ} are satisfied in hydrodynamics -- a theory satisfying conservation laws and the second law of thermodynamics.

Now that we established the equilibrium conditions on coefficients $A$, $D$ and $Q$, we are ready to take on a more intricate case of $N=3$ in Eq.~\eqref{eq:u.dW3} which involves nontrivial nonlinear properties of hydrodynamic equations. Again, setting all gradients to zero in Eq.~\eqref{eq:u.dW3} we find for the terms linear in $q$:
\begin{multline}
    0=3i\left[  q_1\cdot  A_{a_1}^{~b}\Weq_{ba_2a_3}
    -2  q_2\cdot A_{a_1}^{~bc} \Weq_{ba_2}\Weq_{ca_3}
    \right]_{\overline{123}}
    = 3i\left[  q_1\cdot  A_{a_1}^{~b}\Weq_{ba_2a_3}
    -2  q_1\cdot A_{a_2}^{~bc} \Weq_{ba_1}\Weq_{ca_3}
    \right]_{\overline{123}}\\
=  i q_1\cdot\left\{\left[ A_{a_1}^{~b}\Weq_{ba_2a_3}
-\left( A_{a_2}^{~bc}\Weq_{ba_1}\Weq_{ca_3} + (2\leftrightarrow3)\right)\right]-(1\leftrightarrow3)\right\}
+ (1\leftrightarrow2)
\label{eq:0qA3W}
\end{multline}
where we eliminated  $q_3$ by expressing it via $ q_3=- q_1- q_2$. 
Since $q_1$ and $q_2$ are independent vectors, the coefficients of each in the above equation must vanish.

Using Eq.~\eqref{eq:A3A'} and the definition of $\bm\Omega$, Eq.~\eqref{eq:Omega-def}, we can write $\bm A_{a_2}^{~bc}\Weq_{ba_1}=\bm\Omega_{a_2a_1}^{~~~,c}+\bm A_{a_2}^{~b}\Weq_{ba_1}^{~~,c}$. Substituting this into Eq.~\eqref{eq:0qA3W} we find
\begin{equation}
    0=-\bm\Omega_{a_2a_1}^{~~~,c}\Weq_{ca_3} + \bm A_{a_1}^{~b} \Weq_{ba_3/a_2} - \bm A_{a_2}^{~b} \Weq_{ba_1/a_3} - (1\leftrightarrow3)\,,
\label{eq:OmegaAW}
\end{equation}
where we introduced 
\begin{equation}
    \Weq_{ab/c}\equiv \Weq_{abc} - \Weq_{ab}^{~~,d}\Weq_{dc}\,.
\label{eq:W2/1=W3-WW}
\end{equation}

Finally, for the dissipation and noise terms in Eq.~\eqref{eq:u.dW3} in equilibrium we find,
\begin{multline}
    \label{eq:0DWQ3}
    0 = 3\bigg\{
  - D_{a_1}^{~b}\args{q_{1}, q_{1}}
   \Weq_{ba_2a_3}
  -  2 
    \left[ D_{(3)} \args{q_{2},q_{2}} + {\Dt}_{(3)} \args{q_{3}, q_{2}}\right]_{a_1}^{~bc}\,
    \Weq_{ba_2}\Weq_{ca_3}
    \\
    +2\left[Q_{(3)}\args{q_2,q_2}+{\Qt}_{(3)}\args{q_3,q_2}\right]_{a_1a_2}^{~~~~c}\Weq_{c a_3}\bigg\}_{\overline{123}}\,.
\end{multline}
where we suppressed Lorentz indices again, using the notation for bilinear operators, as in Eq.~\eqref{eq:Dqq}.
We can express ``diagonal'' bilinears such as $(q_1,q_1)$ in terms of the ``off-diagonal'' ones, such as $(q_3,q_1)$ using the fact that $q_1+q_2+q_3=0$. E.g., $(q_1,q_1)=-(q_3,q_1)-(q_2,q_1)$. Under the permutation average $\overline{123}$ we can then permute indices until all terms are written in terms of the same off-diagonal bilinear. We choose that bilinear to be $(q_3,q_2)$ because two terms already have this form.
Using the definitions in Eq.~\eqref{eq:DWQ} and Eq.~\eqref{eq:W2/1=W3-WW}  we find
\begin{equation}\label{eq:D-DQ-Q}
\Big[D_{a_3}^{~b}\Weq_{ba_2/a_1}+E_{a_3a_2}^{~~~~,c}\Weq_{ca_1}+\Big(\big(
    D_{(3)}-\Dt_{(3)}\big)_{a_1}^{~~bc}\Weq_{ba_2}
    -\big(Q_{(3)} - \tilde Q_{(3)}\big)_{a_1a_2}^{~~~~c}\Big)\Weq_{ca_3}
    \Big]\args{q_3,q_2}
    \bigg|_{\overline{123}}=0\,.
\end{equation}

We shall now show that also conditions necessary for equilibrium solution to exist in the three-point evolution equations given by Eqs.~\eqref{eq:OmegaAW} and \eqref{eq:D-DQ-Q} are fulfilled in hydrodynamics as a consequence of conservation laws and the second law of hydrodynamics. 

\section{The second law of thermodynamics and equilibrium}
\label{sec:hydro-conserved}

The purpose of this section is to show that, in hydrodynamics, the equilibrium conditions derived in the previous section follow from the two defining properties of hydrodynamics. The first property is that the variables used in hydrodynamic equations are {\em conserved} densities. The second is that the evolution equations satisfy the second law of thermodynamics --- the law of non-decreasing entropy. To focus on these properties, we shall present hydrodynamic equations in the most general multi-component form, without explicitly specifying the variables other than saying that they are {\em conserved} densities. A specific example of such quantities, which we shall consider in more detail in Section~\ref{sec:ABCD-hydro}, are the energy and momentum in a given frame, e.g., the variables defined in Section \ref{sec:ave-Landau}.

Conserved quantities play the central role in {\em thermodynamics}. The thermodynamic state of a system is characterized by the values of such quantities, since they remain constant over time in a closed system. The key quantity in thermodynamics is the entropy --- a logarithm of the density of states, i.e., of the number of microscopic states with given values of conserved quantities. The dependence of the entropy on the values of the conserved quantities --- the equation of state --- contains the information about the microscopic constitution of the theory needed for thermodynamic description.

In hydrodynamics we consider {\em local} thermodynamic equilibrium and the natural dynamic variables are spatial densities of the conserved quantities characterizing the (local) state of thermodynamic equilibrium. We shall denote the set of such variables as $\psi_a$, as we have done already. The entropy density $s(\{\psi_a\})$ is a function of such variables. We define conjugate variables $\chi^a$ as  follows:
\begin{equation}
    ds=\chi^a d\psi_a\,,
 \mbox{ i.e. }
 \chi^a\equiv\frac{\partial s}{\partial\psi_a}\equiv s^{,a}\,.
 \label{eq:ds}
\end{equation}
This relation is, essentially, the first law of thermodynamics.

In equilibrium, the fluctuations of the conserved densities are described by the probability distribution functional $e^S$, where $S=\int d^3x(s-\bar\chi^a\psi_a)$, where $\bar\chi^a$ are parameters (such as inverse temperature) reflecting global conservation laws and which act as Lagrange multipliers controlling equilibrium averaged values of conserved quantities $\Psi_a\equiv\int dx^3\psi_a$. The terms $ \bar\chi^a\psi_a$, being linear in $\psi$, do not affect the second and higher derivatives of $S$ determining fluctuations and correlations.

When we consider fluctuations averaged over a small but macroscopic volume $V$, the central limit theorem dictates that the fluctuations are small and the saddle point (Gaussian) approximation is justified in the calculation of correlation functions in thermodynamic limit $V\to\infty$. We can then find the equilibrium value of the two-point correlation function $\langle\delta\phi_a\delta\phi_b\rangle=-\delta^2S/\delta\psi_a\delta\psi_b$, whose Wigner transform $\Weq_{ab}$ is thus given by the inverse of the matrix $-s^{,ab}$, i.e.,
\begin{equation}\label{eq:SW=delta}
    s^{,ab}\,\Weq_{bc}=-\delta^a_c\,.
\end{equation}
The saddle-point calculation of the three-point function leads to
\begin{equation}
    \Weq_{abc}=\Weq_{ad}\Weq_{be}\Weq_{cf}s^{,def}\,,
    \label{eq:W3-S3}
\end{equation}
which is a familiar result represented by a tree-level (unamputated) diagram.

Taking derivative of Eq.~\eqref{eq:SW=delta} with respect to $\psi$ and using Eq.~\eqref{eq:W3-S3} we find
\begin{equation}
    \Weq_{abc}=\Weq_{ab}^{~~,d}\Weq_{dc}\,, \quad\mbox{i.e.,}\quad \Weq_{ab/c}=0\,.
\label{eq:W3=W'W}
\end{equation}
Therefore, the terms containing $\Weq_{ab/c}$ in the equilibrium conditions given by Eq.~\eqref{eq:OmegaAW} and Eq.~\eqref{eq:D-DQ-Q} vanish.

In order to see the cancellation of the remaining terms we need to consider hydrodynamic equations. We shall do so in a very general form in order to emphasize that this cancellation is not accidental and the properties responsible for it are the conservation laws as well as the second law of thermodynamics, as our general physical understanding of thermodynamic equilibrium tells us.

Since variables $\psi_a$ are {\em conserved} densities their evolution is governed by conservation equations:
\begin{equation}\label{eq:dtpsi=dJ}
    \partial_t\psi_a= - \bm\nabla\cdot\bm j_{a}\,.
\end{equation}
To close these equations we need to express the currents $\bm j_a$ in terms of the variables $\psi_a$ and their derivatives using so-called constitutive equations. The form of the constitutive equations is strongly constrained by the second law of thermodynamics:
\begin{equation}
    \partial_t s  +\bm\nabla\cdot\bm s\ge0\,.
\end{equation}

\subsection{Ideal hydrodynamics}

At the lowest order in derivatives (the zeroth order) there is no dissipation and the second law amounts to the condition that the entropy is conserved, i.e., entropy density also obeys conservation equation:
\begin{equation}
    \partial_t s = -\bm\nabla\cdot\bm s \quad\mbox{(ideal)}\,.
\end{equation}
Using the first law of thermodynamics given by Eq.~\eqref{eq:ds} and differentiating by parts we can write
\begin{equation}
    \partial_t s = -\bm\nabla\cdot (\chi^a\bm j_a) + \bm j_a \bm\nabla\chi^a\,.
\label{eq:dts}
\end{equation}
In order for the last term to be a total divergence, $\bm j_a$ must be a derivative w.r.t. $\chi^a$, i.e.,
\begin{equation}
    \bm j_a = \frac{\partial\bm j}{\partial\chi^a}\equiv \bm j_{,a}\quad\mbox{(ideal)}\,,
\label{eq:ja-ideal}\end{equation}
for some vector function $\bm j$. We distinguish $\chi^a$ derivatives here from $\psi_a$ derivatives, Eq.~\eqref{eq:A3A'}, by the index position (subscript vs superscript).

Thus, the second law of thermodynamics requires that ideal hydrodynamic equations have the form
\begin{equation}
    \partial_t\psi_a=-\bm\nabla\cdot\bm j_{,a} = -\bm j_{,ab}\cdot\bm\nabla\chi^b = -\bm j_{,ab}s^{,bc}\bm\nabla\psi_c\quad\mbox{(ideal)}\,,
\end{equation}
where we used Eq.~\eqref{eq:ds}.
Comparing to Eq.~\eqref{eq:u.dpsi} we identify the vector valued matrix $\bm A\equiv\bm e\cdot A$:
\begin{equation}
    \bm A_a^{~c}=-\bm j_{,ab}s^{,bc}\,.
\label{eq:Ajs}
\end{equation}
Using Eq.~\eqref{eq:SW=delta} we then find that the matrix $\bm\Omega$ defined in Eq.~\eqref{eq:Omega-def} 
is given by
\begin{equation}
    \bm\Omega_{ab}=\bm j_{,ab}\,.
\end{equation}
It is obviously symmetric  due to the commutativity of partial derivatives, which fulfills the equilibrium condition in Eq.~\eqref{eq:Omega12}.

Similarly, the expression appearing in Eq.~\eqref{eq:OmegaAW}
\begin{equation}
    -\bm\Omega_{a_2a_1}^{~~~,c}\Weq_{ca_3} = 
    (\bm j_{,a_2a_1})_{,a_3}=\bm j_{,a_1a_2a_3}
\end{equation}
is fully $(123)$ symmetric and thus the first term in Eq.~\eqref{eq:OmegaAW} also cancels.

Consequently, we see that in hydrodynamics, owing to the second law, both equilibrium conditions in Eqs.~\eqref{eq:Omega12} and \eqref{eq:OmegaAW} are fulfilled.

\subsection{Dissipative hydrodynamics}

In order to check the equilibrium conditions involving $D$ and $Q$ coefficients we need to consider hydrodynamics with noise and dissipation.
The second law of thermodynamics constrains the form of dissipative and noise contributions, $\Delta \bm j_a$, to the current $\bm j_a$:
\begin{equation}
    \bm j_a=\bm j_{,a}+\Delta \bm j_a\,,
\end{equation}
 The local entropy production rate is given by
\begin{equation}
    \partial_\mu s^\mu =\partial_t s+\bm\nabla\cdot\bm s = \Delta j^\mu_a\partial_\mu\chi^a\,,
\label{eq:d.s}
\end{equation}
according to
Eq.~\eqref{eq:dts}, 
where
\begin{equation}
    \bm s = \chi^a \bm j_a - \bm j\,,
\label{eq:s-j}\end{equation}
and we defined $j^\mu_a\equiv \bm e^\mu\cdot\bm j_a$.

The second law of thermodynamics, i.e., the semi-positivity of the  entropy production rate $\partial\cdot s$ in Eq.~\eqref{eq:d.s}, is satisfied by
\begin{equation}
    \Delta j_a^\mu = \lambda^{\mu\nu}_{ab}\partial_\nu\chi^b + \eta^{\widehat c}\sigma^{\mu}_{\widehat c a}\,.
\end{equation}
Fluctuation-dissipation theorem relates the dissipation coefficients $\lambda^{\mu\nu}_{ab}$ to the noise coefficients $\sigma^{\mu}_{\widehat c a}$:
\begin{equation}
    \lambda^{\mu\nu}_{ab}=\delta^{\widehat c\widehat d}\sigma^\mu_{\widehat c a}\sigma^\nu_{\widehat d b}\,,
\label{eq:lambda-sigma}
\end{equation}
The local entropy production rate is given by, according to Eq.~\eqref{eq:dts}, 
\begin{equation}
\partial_\mu s^\mu=\lambda^{\mu\nu}_{ab}(\partial_\mu\chi^a)(\partial_\nu\chi^b) = \delta^{\widehat c\widehat d}(\sigma_{\widehat c a}^\mu\partial_\mu\chi^a)(\sigma_{\widehat d b}^\nu\partial_\nu\chi^b)\ge0\,.
\end{equation}

Substituting current $j^\mu_a$ into Eq.~\eqref{eq:dtpsi=dJ}
(with stochastic variable $\tpsi$ instead of $\psi$) 
\begin{equation}
    \partial_t\tpsi_a + \bm\nabla\cdot\bm j_{,a}=-\partial_\mu\Delta j^\mu_a = 
    -\lambda^{\mu\nu}_{ad}s^{,db}\partial_\mu\partial_\nu\tpsi_b - (\lambda^{\mu\nu}_{ad}s^{,db})^{,c}\partial_\mu\tpsi_c\partial_\nu\tpsi_b + \partial_\mu\eta^{\widehat c}\sigma_{\widehat ca}^\mu + \eta^{\widehat c}\sigma_{\widehat ca}^{\mu~,b}\partial_\mu\tpsi_b\,.
\label{eq:dtpsi-lambda-sigma}
\end{equation}
and comparing the first and the third terms to Eq.~\eqref{eq:u.dpsi} we identify
\begin{equation}
    D_{~~a}^{\mu\nu~b} = -\lambda^{\mu\nu}_{ad}s^{,db}\,;
    \quad
    C_{\widehat ca}^\mu = \sigma_{\widehat ca}^\mu\,,
\label{eq:D-lambda}
\end{equation}
From Eqs.~\eqref{eq:lambda-sigma} and \eqref{eq:Q} we find
\begin{equation}
    Q^{\mu\nu}_{ab}=\lambda^{\mu\nu}_{ab}\,.
\end{equation}
Substituting this equation together with Eq.~\eqref{eq:D-lambda} into Eq.~\eqref{eq:DWQ} and using \eqref{eq:SW=delta} we find that 
\begin{equation}
    E^{\mu\nu}_{ab}\equiv(D\Weq-Q)^{\mu\nu}_{ab}=0\,.
\label{eq:E=0}
\end{equation}
Thus the equilibrium condition in Eq.~\eqref{eq:DWQ} is fulfilled.

From the remaining terms in Eq.~\eqref{eq:dtpsi-lambda-sigma} we identify
\begin{equation}
    \Dt^{\mu\nu~bc}_{~~a}=
    -  
    (\lambda^{\mu\nu}_{ad}s^{,db})^{,c}
    \,;
    \quad
    \Ct_{\widehat ca}^{\mu~c}=C_{\widehat ca}^{\mu~,c}\,.
\label{eq:Dt-Ct}
\end{equation}
Using Eqs.~\eqref{eq:D-lambda}, \eqref{eq:Dt-Ct}, \eqref{eq:Q3C'C}, 
and \eqref{eq:BDC'}
we find
\begin{equation}
   \Dt^{\mu\nu~bc}_{~~a}=D^{\mu\nu~bc}_{~~a}
   \quad\mbox{and}\quad \Qt_{ab}^{\mu\nu~c}=Q_{ab}^{\mu\nu~c}\,.
\label{eq:Qt-Q}
\end{equation}
Eqs.~\eqref{eq:Qt-Q} together with Eqs. \eqref{eq:W2/1=W3-WW} and \eqref{eq:E=0} mean that the equilibrium condition in Eq.~\eqref{eq:D-DQ-Q}
is satisfied.
 
We conclude that equilibrium conditions for two-point (Eqs.~\eqref{eq:Omega12} and \eqref{eq:DWQ}) and three-point (Eqs. \eqref{eq:OmegaAW}, \eqref{eq:D-DQ-Q}) correlators found in the previous section  are satisfied in hydrodynamics as a consequence of the conservation equations and the second law of thermodynamics.\footnote{It is also possible to show that upon general {\em nonlinear} transformation of variables in Eq.~\eqref{eq:u.dpsi}, which alters coefficients $ABCD$, the resulting coefficients still satisfy the equilibrium conditions in Section~\ref{sec:equilibrium}, as long as the original coefficients came from hydrodynamics in terms of primary variables. An example of such a transformation is the change of variables from $\teps$ and $\tn$ to pressure and specific entropy, as in Ref.~\cite{An_2023}.}

\subsection{Simpler evolution equations}
\label{sec:evolution-conserved}

The general equations we derived for the correlators simplify to a significantly more intuitive form when the variables are chosen to be conserved densities. In addition to Eqs.~\eqref{eq:Qt-Q} which we found helpful for verifying the equilibrium conditions, there is also another property useful for simplifying correlator evolution equations. Using Eq.~\eqref{eq:Ajs} one can show the symmetry of $A_a^{~bc}$ with respect to $bc$, which is not true generally:
\begin{equation}
    A_a^{~bc}=A_a^{~b,c}=-(\bm j_{,ad}s^{,db})^{,c}=-\bm j_{,ade}s^{,ec}s^{,db}-\bm j_{,ad}s^{,dbc}=A_a^{~cb}\,.
\label{eq:A3-sym}
\end{equation}

With this equation and Eqs.~\eqref{eq:Qt-Q} we can simplify Eqs.~\eqref{eq:W3-eqs} for three-point Wigner functions as follows. Instead of Eq.~\eqref{eq:Ah3W}, operator $\Ah_a^{bc}$, or rather its symmetrized form appearing in Eq.~\eqref{eq:u.dW3}, can be now written as 
\begin{equation}
\Ah_{a_1}^{~bc} \args{W_{ba_2}^{(-q_2',q_2')}
   ,W_{ca_3}^{(-q_3',q_3')}}\Big|_{\overline{23}}
= \Ah_{a_1}^{~b,c}\left(W_{ba_2}^{(-q_2',q_2')}
   W_{ca_3}^{(-q_3',q_3')}\right)\,,
\end{equation}
where $\Ah_a^{~b,c}$ is the derivative of the operator $\Ah_a^{~b}$ already defined in Eq.~\eqref{eq:Ah} 
for $N=2$ equation. Note that, due to Eqs.~\eqref{eq:A3-sym} and \eqref{eq:Qt-Q}, $\Ah_a^{b,c}=\Ah_a^{c,b}$. Now, unlike Eq.~\eqref{eq:Ah3W}, the operator simply acts on a product of two Wigner functions instead of a bilinear.
The wave-vector in Eq.~\eqref{eq:Ah} is $q_1=-q_2-q_3$ and the derivative $\cfd$ obeys the usual product rule, $\cfd(AB)=(\cfd A)B+A(\cfd B)$. The derivative $\partial/\partial q_1$ should be understood as $-\partial/\partial q_2-\partial/\partial q_3$ (in accordance with $\sum_i\partial/\partial q_i=0$).

Similarly, Eq.~\eqref{eq:VQ3W} simplifies to
\begin{equation}
    \VQW_{a_1a_2}^{~~~b}W_{ba_3}^{(-q_3,q_3)}\Big|_{\overline{12}}\equiv-2
    Q_{a_1a_2}^{~~~,b}(q_1,q_2)
    W_{ba_3}^{(-q_3,q_3)}\Big|_{\overline{12}}=\VQW_{a_1a_2}^{~~~,b}W_{ba_3}^{(-q_3,q_3)}\Big|_{\overline{12}}\,,
\end{equation}
where, as before, $\VQW_{a_1a_2}$ is already defined in Eq.~\eqref{eq:VQqq} for $N=2$ equation. 

With everything put together, in a more explicit form, Eq.~\eqref{eq:W2-eqs} for $N=2$ reads 
\begin{multline}\label{eq:W2-cons}
    u\cdot\cfd W_{a_1a_2}^{(q_1,q_2)}=2\Bigg[
\left(A\cdot\left(iq_{1} + \frac12 \cfd\right)- \frac{\partial}{\partial q_{1\lambda}}\KK_\lambda^\mu q_{1\mu} 
+ B - D(q_1,q_1) \right)_{a_1}^{~b}W_{ba_2}^{(q_1,q_2)}
-Q_{a_1a_2}(q_1,q_2)
    \Bigg]_{\overline{12}}\,,
\end{multline}
with matrix $K$, defined as in Eq.~\eqref{eq:DeltaA},
\begin{equation}\label{eq:K4}
    \KK_\lambda^\mu\equiv 
   \cfd_\lambda A^\mu
   -\partial_\lambda u^\mu
  + (a_\lambda+2\omega_{\lambda\nu}A^\nu)A^\mu\,.
\end{equation}

Similarly, for $N=3$, if the variables are chosen to be conserved densities (primary variables in our terminology), then, due to Eqs.~\eqref{eq:Qt-Q} and~\eqref{eq:A3-sym}, we can write Eqs.~\eqref{eq:W3-eqs} more explicitly in the form
\begin{multline}\label{eq:W3-cons}
    u\cdot\cfd W_{a_1a_2a_3}^{(q_1,q_2,q_3)} = 3\Bigg[
    \left(A\cdot\left(iq_{1} + \frac13 \cfd\right)- \frac{\partial}{\partial q_{1\lambda}}\KK_\lambda^\mu q_{1\mu}  
    + B - D(q_1,q_1)\right)_{a_1}^{~b}W_{ba_2a_3}^{(q_1,q_2,q_3)}
    \\
    +\left[\left(A\cdot\left(iq_{1} + \frac12 \cfd\right)- \frac{\partial}{\partial q_{1\lambda}}\KK_\lambda^\mu q_{1\mu} 
    + B - D(q_1,q_1)
    \right)_{a_1}^{~b,c} + \hat\Pi_{a_1}^{~bc}\right]\left(W_{ba_2}^{(-q_2',q_2')}W_{ca_3}^{(-q_3',q_3')}\right)
    \\ -2 Q_{a_1a_2}^{~~~,c}(q_1,q_2)W_{ca_3}^{(-q_3,q_3)}
    \Bigg]_{\overline{123}}\,.
\end{multline}

As a consistency check we could apply these equations for the far simpler case of a single conserved density variable, i.e., the problem of diffusion studied in Ref.~\cite{An_2021}. The matrices $A$, $B$ are absent, because there is no flow, while the matrices $D$ and $Q$ become simply numbers, $D\to\gamma$ and $Q\to\lambda$, thus reducing Eqs.~\eqref{eq:W2-cons}, and \eqref{eq:W3-cons} to the corresponding equations for $\WN$ in Ref.~\cite{An_2021}.

\section{$ABCD$s of relativistic hydrodynamics}
\label{sec:ABCD-hydro}

In this Section we demonstrate how to write manifestly confluent and SO(3) covariant equations for fully {\em nonlinear} {\em stochastic} hydrodynamics and determine the coefficients $ABCD$ and $Q$ needed for the deterministic  correlator evolution equations.

Let us first translate the compact and general, but abstract notation of the previous section into a more explicit and recognizable notation. The components of the set of variables are given by
\begin{equation}\label{eq:tpsi}
    \tpsi_a=\{\bm\pi,\teps, \tilde n\}\,,
\end{equation}
corresponding to the momentum density, energy density, and a charge density. For now, tilde over the symbols can be ignored, but it will play a role when we consider these densities in two different frames. 

From the first law of thermodynamics,
\begin{equation}
    d\tilde s=-{\bm\beta}\cdot d\bm\pi+\tilde\beta d\teps -\alpha d\tn\,,
\label{eq:d-tilde-s}
\end{equation}
we read off the set of conjugate variables defined in Eq.~\eqref{eq:ds}:
\begin{equation}
    \chi^a=\{\chi^{\bm\pi},\chi^\teps,\chi^\tn\}=\{-{\bm\beta},\tilde\beta,-\alpha\}\,.
\label{eq:chi-beta}
\end{equation}
The Legendre transform of $\tilde s$ is $\tilde\beta p$, where $p$ is pressure:
\begin{equation}
 \tilde \beta p = \tilde s- \chi^a\psi_a=\tilde s -\tilde\beta\teps + \alpha \tn + \bm\beta\cdot\bm\pi\,.
\label{eq:pressure}
\end{equation}
Therefore, $(\tilde\beta p)_{,a}=-\psi_a$ and derivatives of pressure are, thus,
\begin{equation}
    \tilde\beta p_{,a}=-\{\bm\pi,\tilde w,\tn\}\,,
\quad \mbox{
where $\tilde w\equiv \teps+p$}\,.
\label{eq:tbetap}
\end{equation}
Vector $\bm j$ introduced in the previous section is given by
\begin{equation}
    \bm j = p{\bm\chi}=-p\bm\beta\,.
\end{equation}
The ideal parts of the currents $\bm j_a$ carrying the conserved densities are thus given by
\begin{equation}\label{eq:j,a}
    \bm j_{,a}=p_{,a}\bm\chi+p\bm e_a = \{\pi_a\bm v + p\bm e_a,\tilde w\bm v,\tn\bm v\}\,,
\end{equation}
where we defined
\begin{equation}
    \bm v =  -\frac{\chi^{\bm\pi}}{\chi^\teps} = \frac{\bm\beta}{\tilde\beta}\,.
\label{eq:v-beta}
\end{equation}
Substituting Eqs.~\eqref{eq:tpsi} and~\eqref{eq:ja-ideal} into Eq.~\eqref{eq:dtpsi=dJ} we find the familiar set of ideal hydrodynamic equations:
\begin{equation}
\partial_t\pi_a =  - \bm\nabla\left(\pi_a\bm v\right)-\nabla_a p\,;\quad \partial_t\teps=-\bm\nabla(\tilde w\bm v)\,;\quad \partial_t \tn = -\bm\nabla(\tn\bm v)\quad \mbox{(ideal)}\,. 
\end{equation}
The ideal part of the entropy current, Eq.~\eqref{eq:s-j}, is given by
\begin{equation}
    \bm s = \chi^a\bm j_{,a} - \bm j = \chi^a p_{,a}\bm\chi=s\bm v \quad \mbox{(ideal)}\,.
\end{equation}

So far we have not assumed Lorentz invariance. Now, in a Lorentz invariant theory $(\tilde\beta,\bm\beta)$ transform as components of a four-vector, while $\alpha$ and $p$ are scalars.  Therefore, pressure is a function of two scalar variables:~$\alpha$ and
\begin{equation}
\breve\beta\equiv\sqrt{\tilde\beta^2-\bm\beta^2}=\tilde\beta/\gamma\,,
\label{eq:beta-beta}
\end{equation}
with $\gamma\equiv1/\sqrt{1-\bm v^2}$, where we used Eq.~\eqref{eq:v-beta}. The derivatives of pressure $p(\breve\beta,\alpha)$ are given by
\begin{equation}
    \breve\beta dp=-\breve w d\breve\beta +\breve n d\alpha\,;\quad\mbox{with}
    \quad \breve w=\breve\varepsilon + p\,.
\label{eq:bbetap}
\end{equation}
Obviously, $\bm v$ is the velocity of the frame were $\bm\beta$ vanishes --- the rest frame of the fluid, while $\breve\varepsilon$ is the rest frame energy density, $\breve T=1/\breve\beta$ is the rest frame (proper) temperature and $\breve \mu=\alpha/\breve\beta$ is the rest frame chemical potential.

Since the function $p(\breve\beta,\alpha)$ is the equation of state (determined, e.g., by lattice gauge theory calculation) serving as an input for hydrodynamics, we would like to express the $ABCDQ$ coefficients in terms of that function and its derivatives, together with the velocity $\bm v$.

The relationships between the two sets of variables, $\tpsi=\{\bm\pi,\teps,\tn\}$ and $\{\bm v,\breve\varepsilon,\breve n\}$ can be obtained by differentiating $p(\breve\beta,\alpha)=p\left(\sqrt{\tilde\beta^2-\bm\beta^2},\alpha\right)$ and using Eqs.~\eqref{eq:tbetap}, \eqref{eq:bbetap}. We thus find
\begin{equation}
    \bm\pi=\breve w\gamma^2\bm v\,,\quad \tilde w = \breve w\gamma^2\,,\quad
    \tilde n = \breve n\gamma\,.
\label{eq:pivwn}
\end{equation}
Combining these equations we can write more relationships which will be also useful:
\begin{equation}
    \bm\pi=\tilde w\bm v\,,\quad \teps = \breve\varepsilon + \bm\pi\cdot\bm v\,.
\label{eq:piv}
\end{equation}

The rest frame entropy density $\breve s$ is related to pressure by Legendre transformation as usual:
\begin{equation}
    \breve s = \breve\beta p + \breve\beta\breve\varepsilon -\alpha\breve n=\tilde s/\gamma
    \,.
\end{equation}

We can also rewrite conservation laws in Eq.~\eqref{eq:dtpsi=dJ} in a manifestly Lorentz covariant form by expressing $\bTmunu$ and $\breve J^\mu$ in terms of the variables $\tpsi_a$ and their currents $j_a^\mu$:
\begin{equation}
    \bTmunu = (\teps u_\nu+\pi_\nu)u^\mu + j_\teps^\mu u^\nu + j_{\pi_\nu}^\mu = \tilde w (u+ v)^\mu(u+ v)_\nu + \delta^\mu_\nu p + \Delta\bTmunu\,;
\end{equation}
\begin{equation}
    \breve J^\mu = \tn u^\mu + j_{\tn}^\mu = \tn(u+v)^\mu + \Delta \breve J^\mu\,,
\label{eq:J=}
\end{equation}
where 
\begin{equation}\label{eq:v}
    v^\mu\equiv \bm e^\mu\cdot\bm v =\pi^\mu/\tilde w
\end{equation}
from Eq.~\eqref{eq:piv} and $\Delta\bTmunu$, $\Delta\breve J^\mu$
are dissipative (non-ideal) contributions which are the corresponding components of $\Delta j_a^\mu$ introduced in the previous Section: $\Delta T^\mu_{~\nu}=j^\mu_\varepsilon u_\nu + \Delta j_{\pi_\nu}^\mu$ and $\Delta J^\mu=\Delta j^\mu_\tn$. 

It is straightforward to check that conservation laws applied to ideal parts of $\bTmunu$ and $\breve J^\mu$ imply conservation of (the ideal)  entropy current:
\begin{equation}
    s^\mu = \tilde s(u+v)^\mu 
    \quad\mbox{(ideal)}\,.
\end{equation}
The easiest way to see this is by using
\begin{equation}
    \breve u=\gamma(u+v) 
\end{equation}
and Eqs.~\eqref{eq:pivwn} to rewrite ideal contributions in a more familiar form: 
\begin{equation}
    \bTmunu=\breve w\breve u^\mu\breve u_\nu + \delta^\mu_\nu p\,,\quad \breve J^\mu=\breve n\breve u^\mu\,,\quad\mbox{and}\quad s^\mu = \breve s \breve u^\mu \quad\mbox{(ideal)}\,.
\end{equation}

\subsection{Ideal hydrodynamics --- $AB$}

Let us first concentrate on the ideal part of the hydrodynamic equations, which would give us coefficients $A$ and $B$ in Eq.~\eqref{eq:u.nablapsi-ABCD}.
We can obtain equation for $u\cdot\partial\teps$ from $u^\nu\partial_\mu\bTmunu=0$:
\begin{equation}
    u\cdot\partial\teps = -\tilde w\,\partial\cdot u -\partial\cdot\pi - 
    \pi\cdot a - v\cdot(\pi\cdot\partial)u\,,
\label{eq:u.dteps}
\end{equation}
where $a\equiv (u\cdot\partial)u$ is the average acceleration of the fluid.

To ``confluentize'' this equation we use the confluent derivative defined by Eq.~\eqref{eq:barnabla-def} and Eq.~\eqref{eq:nabla-pi}:
\begin{equation}
    \cfd_\mu\teps = \partial_\mu\teps\,;\quad
    \cfd\cdot\pi = \partial\cdot\pi +\ucon_{\mu\lambda}^\mu\pi^\lambda\,;
\end{equation}
while, from Eq.~\eqref{eq:bar-omega},
\begin{equation}
    \ucon_{\mu\nu}^\mu=u_\nu \partial\cdot u - a_\nu\,.
\label{eq:omega-a}
\end{equation}
Substituting into Eq.~\eqref{eq:u.dteps} we find
\begin{equation}
     u\cdot\cfd\teps = -\tilde w\,\partial\cdot u - \cfd\cdot\pi  - \pi\cdot(v\cdot\partial)u -2\pi\cdot a\,.
\end{equation}
Comparing this to the fully nonlinear equation~\eqref{eq:u.nablapsi-ABCD} we identify the coefficients $A^{\mu~b}_{~a}$ and $B_a$ for $a=\teps$:
\begin{equation}
A^{\mu~\teps}_{~\teps}  =0\,,\quad  A^{\mu~\bm\pi}_{~\teps} = - \bm e^\mu\,, \quad B_\teps= -\tilde w\,\partial\cdot u - \pi\cdot(v\cdot\partial)u  -2\pi\cdot a\,,
\label{eq:A2Beps}
\end{equation}
where we use the confluently constant local Cartesian triad $\bm e^\mu$ orthogonal to $u^\mu$, as in Eqs.~\eqref{eq:nabla-e}, \eqref{eq:ring-o}, to express $\pi^\mu$ in terms of the local confluent Cartesian coordinates $\bm\pi$: $\pi^\mu=\bm e^{\mu}\cdot\bm\pi$.

Now, we repeat this for the $\Delta^\nu_\lambda(u\cdot\partial)\pi_\nu$ equation obtained from $\Delta^\nu_\lambda\partial_\mu \bTmunu=0$:
\begin{equation}
    \Delta^\nu_\lambda(u\cdot\partial)\pi_\nu
    = -\Delta^\nu_\lambda\partial_\nu p 
    - \Delta^\nu_\lambda\partial_\mu\left(v^\mu\pi_\nu\right)
    - \pi_\lambda\,\partial\cdot u
    - (\tilde w u + \pi)\cdot\partial\, u_\lambda\,.
\end{equation}
Using properties of the connection $\ucon$ from Eq.~\eqref{eq:bar-omega}, such as
Eq.~\eqref{eq:omega-a} and $\ucon_{\lambda b}^a\equiv e^a_\mu e_b^\nu\ucon_{\lambda\nu}^\mu=0$, we obtain the confluentized form of this equation:
\begin{equation}
    \Delta^\nu_\lambda(u\cdot\cfd)\pi_\nu
    = -\Delta^\nu_\lambda\cfd_\nu p 
    - \Delta^\nu_\lambda\cfd_\mu\left(v^\mu\pi_\nu\right)
    -\pi_\lambda (v\cdot a)
    - \pi_\lambda\,\partial\cdot u
    - (\tilde w u + \pi)\cdot\partial\, u_\lambda \,.
\end{equation}
In terms of the local confluent Cartesian basis components:
\begin{equation}
    (u\cdot\cfd)\bm\pi
    = -\bm\cfd p 
    - \cfd_\mu\left(v^\mu\bm\pi\right)
    -\bm\pi\left( v\cdot a
    +\partial\cdot u
    \right)
    - \bm e^\lambda\tilde w (u + v)\cdot\partial\, u_\lambda\,,
\end{equation}
where $\bm\cfd\equiv\bm e^\mu\cfd_\mu$.
Comparing this to the fully nonlinear equation~\eqref{eq:u.nablapsi-ABCD} we identify the coefficients $A^{\mu~b}_{~a}$ and $B_a$ for $a=\bm\pi$.
The first two terms produce the $A$ coefficients and the remaining terms --- the $B$ coefficients: 
\begin{subequations}
\begin{align}
    &A^{\mu~\teps}_{~\bm\pi}
    = -p^{,\teps}\bm e^\mu + v^\mu\bm v(1+p^{,\teps})\,;
    \\
    &A^{\mu~\pi_b}_{~\pi_a}
    = - p^{,\pi_b} ( e^{\mu}_a - v^\mu v_a) - (e^{\mu b}v_a + \delta_a^b v^\mu )\,;\label{eq:A2}\\
    &B_{\bm\pi}=-\bm\pi\left( v\cdot a
    +\partial\cdot u
    \right)
    -\tilde w\bm a -\bm e^\lambda(\pi\cdot\partial) u_\lambda
    \,.
\end{align}
\label{eq:A2B1}
\end{subequations}

Coefficients in the equation for charge density $\tn$, $B_\tn$ and $A^{\mu~b}_{~\tn}$ can be obtained similarly from equation $\partial_\mu\breve J^\mu=0$. We shall present them in a section dedicated to hydrodynamics with a conserved charge. Our purpose in this Section is to illustrate the method and, for simplicity, we focus on the equation for the fluid without a conserved charge, where our main challenge --- fluctuations of velocity --- is being met.

In order to write equations for fluctuations, such as Eqs.~\eqref{eq:u.dW2} and \eqref{eq:u.dW3} for Wigner functions, we need coefficients in Eq.~\eqref{eq:u.phi-ABCD} for fluctuations $\phi_a$. Since this equation is obtained by Taylor expanding Eq.~\eqref{eq:u.nablapsi-ABCD} for $\tilde\psi_a$, these coefficients can be calculated by differentiating fully nonlinear coefficients $A$ and $B$, according to Eqs.~\eqref{eq:A3A'},~ \eqref{eq:B2B'}, and~\eqref{eq:BDC'}. For linearized equations we only need matrix $B_a^{~b}$. Its matrix elements can be obtained using Eq.~\eqref{eq:B2B'} from $B_a$ and $A^{\mu~b}_{~a}$ we have just found. We obtain
\begin{subequations}
\begin{align}
 &   B_\teps^{~\teps}=
    -(1+p^{,\teps})\left(
    \partial\cdot u
+v\cdot(v\cdot\partial)u
\right)\,;
\\  
&B_{\teps}^{~\bm\pi}
    = - p^{,\bm\pi}\left(
    \partial\cdot u
-v\cdot(v\cdot\partial)u
\right)- (v^\mu\bm e^\nu+\bm e^\mu v^\nu)\partial_\mu u_\nu
  -2\bm a \,;
\\
 &    B_{\bm\pi}^{~\teps}
    = -(1+p^{,\teps})(\bm a-\bm v\,(v\cdot a))
    -\bm\cfd p^{,\teps}\,;
\\
 &   B_{\pi_a}^{~\pi_b}= -\delta_a^b(v\cdot a+\partial\cdot u)-v_a(a^b-v\cdot a\,p^{,\pi_b})
    -a_a p^{,\pi_b}
    -e^\lambda_a e^{\mu b}\partial_\mu u_\lambda-\cfd_a\,p^{,\pi_b}\,.
\end{align}
   \label{eq:B2}
\end{subequations}

These fully nonlinear expressions can be differentiated further to obtain higher order vertices, such as $A_{(3)}$, $B_{(3)}$, etc.\ according to Eqs.~\eqref{eq:A3A'} and \eqref{eq:BDC'}.

\subsection{Dissipative terms --- $DQ$}

Taking into account dissipative contributions to constitutive equations we find for the divergence of the entropy current constrained to be non-negative by the second law of thermodynamics:
\begin{equation}
  \partial\cdot\left(\breve s\breve u + \Delta\breve s \right) =
    -\Delta\bTmunu \partial_\mu (\breve\beta\breve u^\nu) - \Delta \breve J\cdot\partial\alpha\ge 0\,, 
    \quad\mbox{where}\quad
    \Delta\breve s^\mu = -\breve\beta\breve u^\nu\Delta\bTmunu - \alpha\Delta\breve J^\mu\,.
\label{eq:2ndlaw}\end{equation}

Focusing on the dissipative corrections to energy-momentum tensor here, we conclude that the second law requires
\begin{equation}
    \Delta\bTmunu = - \Mmanb\,\partial_\alpha (\breve\beta\breve u^\beta)\,,
\end{equation}
where $\Mdis$ is a semipositive definite matrix with respect to composite indices $\mu\nu$, $\alpha\beta$: $\Mmanb X_\mu^\nu X_\alpha^\beta \ge0$ for any~$X$. The ``Landau condition'' 
\begin{equation}\label{eq:Landau}
    u_\mu\Delta\bTmunu=0\,,
\end{equation}
 following from Eq.~\eqref{eq:uT-e-pi}, and isotropy further constrain $\Mdis$ down to two coefficients (nontrivial eigenvalues of matrix~$\Mdis$) $\Meta$ and $\Mzeta$ 
\begin{equation}\label{eq:Mmanb}
    \Mmanb =
    {\Meta}\left(\Delta^{\mu\alpha}\Delta_{\nu\beta}+\Delta^\mu_\beta\Delta^\alpha_\nu-\frac23\Delta^\mu_\nu\Delta^\alpha_\beta
    \right)
    + 
    \Mzeta\Delta^\mu_\nu\Delta^\alpha_\beta\,.
\end{equation}
The coefficients are usually expressed in terms of shear $\eta$ and bulk $\zeta$ viscosities:
\begin{equation}    \Meta=\eta\breve\beta^{-1}\ge0 \quad\mbox{and}\quad \Mzeta=\zeta\breve\beta^{-1}\ge0\,.
\end{equation}

Thus, using $\breve\beta\breve u = \tilde\beta(u+v)$ and $\Mmanb u^\beta=0$, we find
\begin{equation}\label{eq:DT=M} 
    \Delta\bTmunu = - \Mmanb\left(
    \tilde\beta\partial_\alpha u^\beta + \partial_\alpha(\tilde\beta  v^\beta)
    \right)\,.
\end{equation}
The first term contributes to averaged equations, but is negligible ($\mathcal O(kq)$) when we consider equations for fluctuations.

Substituting Eq.~\eqref{eq:DT=M} into $u^\nu\partial_\mu\bTmunu=0$ we find (focusing on the $D$ terms only)
\begin{equation}\label{eq:u.de}
    (u\cdot\cfd)\teps = \dots + u^\nu\cfd_\mu(\Delta\bTmunu)\,,
\end{equation}
where, as before, we replaced the partial derivative with the confluent derivative, neglecting $\mathcal O(kq)$ terms, which are beyond our gradient truncation order.
Since $u$ is confluently constant, Eq.~\eqref{eq:barnabla-def}, $u^\nu\cfd_\mu(\Delta\bTmunu)=\cfd_\mu(u^\nu\Delta\bTmunu)=0$ due to the Landau condition in Eq.~\eqref{eq:Landau}.
Thus,
\begin{equation}
    D^{\mu\nu~b}_{~~\teps}=0\,,\quad\Dt^{\mu\nu~bc}_{~~\teps}=0\quad\mbox{for all $b,c\in\{\bm\pi,\teps,\tn\}$}\,.
\end{equation}

Proceeding along similar lines, we find
\begin{equation}
    (u\cdot\cfd)\bm\pi = \dots - \bm e^\nu\cfd_\mu(\Delta\bTmunu) 
    = \dots + \bm e^\nu\cfd_\mu \left[\Mmanb \cfd_\alpha(\tilde\beta v^\beta)
    \right]\,.
\end{equation}
We can then identify:
\begin{equation}
    D^{\mu\nu~b}_{~~a}=\Mdis^{\mu\nu}_{a c}(\tilde\beta v^c)^{,b};    \quad D^{\mu\nu~bc}_{~~\pi_a} =
    \tilde D^{\mu\nu~bc}_{~~\pi_a}
    =D^{\mu\nu~b,c}_{~~\pi_a}\,,
\end{equation}
where
\begin{equation}
    \Mdis^{\mu\nu}_{ac}\equiv e^\alpha_a e^\beta_c \Mdis^{\mu\nu}_{\alpha\beta} 
    \,.
\end{equation}
Note that, since we defined $e^\alpha_a=0$ for $a\ne1,2,3$, $\Mdis^{\mu\nu}_{\teps c}= \Mdis^{\mu\nu}_{a\teps} =0$.

Finally,
\begin{equation} 
    Q^{\mu\nu}_{ab}=\Mdis^{\mu\nu}_{ab} 
    \,. 
\end{equation}

\subsection{Transport of a conserved charge}

We have already written the ideal part of the conserved charge current $\breve J^\mu$ in Eq.~\eqref{eq:J=} in terms of the hydrodynamic variable.
For the dissipative part,
the second law of thermodynamics as expressed by Eq.~\eqref{eq:2ndlaw} constrains $\Delta\breve J$ up to one coefficient 
\begin{equation}
    \Delta \breve J^\mu = - \Msigma\Delta^{\mu\nu}\partial_\nu\alpha\,,
\end{equation}
which is usually expressed in terms of charge conductivity $\sigma$:
\begin{equation}
    \Msigma=\sigma\breve\beta^{-1}\ge0\,.
\end{equation}

To identify the explicit expressions for the $AB$ and $D$ coefficients involving the charge density variable, we write correspoding terms in the current conservation equation in the form of Eq.~\eqref{eq:u.nablapsi-ABCD}:
\begin{equation}
    (u\cdot\cfd)\tn
    =-\cfd\cdot(\tn v)
    -\tn(\partial\cdot u+v\cdot a)+\cfd_\mu( \Msigma\Delta^{\mu\nu}\cfd_\nu\alpha)\,.
\end{equation}
We thus find
\begin{equation}
    B_{\tn}=-\tn(\partial\cdot u+v\cdot a)\,;
    \quad
    A^{\mu~b}_{~\tn} =-(\tn v^\mu)^{,b}\,;
    \quad
    D^{\mu\nu~b}_{~~\tn}=\Delta^{\mu\nu}\Msigma\alpha^{,b}\,,
\end{equation}
where $b\in\{\bm\pi,\teps,\tn\}$. 

\subsection{Variable conversion}

We have seen in this section that the coefficients $ABDQ$ needed to write fluctuation evolution equations can be expressed in terms of hydrodynamic quantities $\{\bm\pi,\teps,\tn\}$, pressure $p$ and velocity $\bm v$ as functions of these variables, and derivatives of these two functions. In practice, we exploit relativistic invariance which allows us to express all this information in terms of a single function --- pressure as a function of scalar variables such as energy and charge densities, or temperature and chemical potential. This is how a theoretic calculation, e.g., lattice QCD calculation, would present to us the equation of state. The purpose of this section is to lay out the framework for obtaining the derivatives with respect to variables such as $\{\bm\pi,\teps,\tn\}$ which we need for $ABDQ$ coefficients in terms of the derivatives of such an equation of state.

If we take the equation of state $p(\breve\varepsilon,\breve n)$ as given, we can then express all derivatives we need 
with respect to $\{\bm\pi,\teps,\tn\}$ using chain rule and the  derivatives of variables $\{\bm v,\breve\varepsilon,\breve n\}$  with respect to $\{\bm\pi,\teps,\tn\}$. These derivatives can be generated starting from explicit expressions for $\{\bm\pi,\teps,\tn\}$ in terms of $\{\bm v,\breve\varepsilon,\breve n\}$ already contained in Eqs.~\eqref{eq:pivwn} and \eqref{eq:piv}. More explicitly, differentiating
\begin{equation}\label{eq:tilde-breve}
\bm\pi = \breve w \gamma^2 \bm v\,,
\quad
\teps=\breve\varepsilon + \breve w \gamma^2 v^2 = (\breve\varepsilon + v^2 p)\gamma^2\,,\quad \tn=\breve n\gamma\,.
\end{equation}
 with respect to 
$\{\bm v,\breve\varepsilon,\breve n\}$ and inverting the resulting  matrix we find
\begin{equation}\label{eq:ven-deriv-2} 
\begin{array}{l@{\qquad}l@{\qquad}l}
        v^{a,\pi_b}
         =\dfrac1{\tilde w}\left(\delta^{ab}+v^av^b\dfrac{p_\varepsilon+c_s^2}{1-v^2c_s^2}\right);& \bm v^{,\teps}=-\dfrac{\bm v}{\tilde w}\dfrac{1+p_\varepsilon}{1-v^2c_s^2}
         \,;&\bm v^{,\tn}=-\dfrac{\bm v}{\gamma\tilde w}\dfrac{p_n}{1-v^2c_s^2}\,
         ;\\[1em]
          \breve\varepsilon^{,\bm\pi}
          =-\bm v\dfrac{2-v^2(c_s^2-p_\varepsilon)}{1-v^2c_s^2}
         \,;&  \breve\varepsilon^{,\teps}=1+v^2\dfrac{1+p_\varepsilon}{1-v^2c_s^2}
         \,;& \breve\varepsilon^{,\tn}=\dfrac{v^2}{\gamma}\dfrac{p_n}{1-v^2c_s^2}
         \,;\\[1em]
         \breve n^{,\bm\pi}=-\bm v\dfrac{\breve n}{\breve w}\dfrac{1+v^2p_\varepsilon}{1-v^2c_s^2}
         \,;&  \breve n^{,\teps}=v^2\dfrac{\breve n}{\breve w}\dfrac{1+p_\varepsilon}{1-v^2c_s^2}
         \,;&\breve n^{,\tn}=\dfrac1\gamma\dfrac{1-v^2p_\varepsilon}{1-v^2c_s^2}
         \,.
    \end{array}
\end{equation}
where
\begin{equation}
    p_\varepsilon\equiv\partial p(\breve\varepsilon,\breve n)/\partial\breve\varepsilon\equiv p^{,\breve\varepsilon}\,,\quad p_n\equiv\partial p(\breve\varepsilon,\breve n)/\partial\breve n\equiv p^{,\breve n}\,,\quad\mbox{and}\quad c_s^2\equiv p_\varepsilon + \breve n p_n/\breve w\,.
\end{equation} Also convenient are these relations:
\begin{equation}
    p^{,\bm\pi}
    =-\bm v\frac{p_\varepsilon+c_s^2}{1-v^2c_s^2}\,;\qquad
    p^{,\teps}=\frac{p_\varepsilon+v^2c_s^2}{1-v^2c_s^2}\,;\qquad
    p^{,\tn}=\frac1\gamma\frac{p_n}{1-v^2c_s^2}\,.
\end{equation}

To calculate derivatives of $\tilde\beta$ needed for $D$ and $Q$ terms, taking equation of state function $\tilde\beta(\breve\varepsilon,\breve n)$ as given, one simply uses the relation $\tilde\beta=\breve\beta\gamma$ from Eq.~\eqref{eq:beta-beta} together with the chain rule.

\section{Phonon kinetics}
\label{sec:phonon}

The purpose of this section is to elucidate the physical significance of the operator $\VABD$, defined in Eqs.~\eqref{eq:H-2pt} or, more precisely, its Wigner transform given by Eqs.~\eqref{eq:W2-eqs}, which is a major building block of the fluctuation evolution equations. In particular, we shall show that, when restricted to the sector of longitudinal (sound) fluctuations, this operator matches the Liouville operator in a kinetic equation describing a gas of phonons propagating in an inhomogeneous and dynamic fluid. More specifically, the matrix $q_\mu K^\mu_\lambda$, Eq.~\eqref{eq:K4}, matches all inertial forces affecting the motion of the phonon in such a fluid, while $D$ matches the sound attenuation coefficient. A similar observation has been made already in Ref.~\cite{An:2019rhf}. Here we wish to demonstrate this in the novel formalism used in this paper. This will also serve as a (very) nontrivial check of our results.

For this check we are going to focus on hydrodynamics without a conserved charge, for simplicity.
For two-point functions we need linearized equations so we will evaluate the nonlinear coefficients in Eqs.~\eqref{eq:A2Beps}, \eqref{eq:A2B1}, and \eqref{eq:B2} setting fluctuations $\phi_a=\{\pi_a,\teps\}$ in them to zero: $\teps\to\varepsilon$ and $v\to0$:
\begin{equation}\label{eq:AB-linear}
    A^{\mu~b}_{~a}=-\begin{pmatrix}
        0 & c_s^2 e^\mu_a\\
        e^{\mu b} & 0
    \end{pmatrix}\,;
   \quad B_a^{~b}=-\begin{pmatrix}
\delta_a^b\,\partial\cdot u + e^{\mu b}e^\nu_a\partial_\mu u_\nu & (1+c_s^2)a_a+\cfd_a(c_s^2)\\
2 a^b & (1+c_s^2)\,\partial\cdot u
    \end{pmatrix}
    \,;
\end{equation}
\begin{equation}
    Q^{\mu\nu}_{ab}=\begin{pmatrix}
        0&0\\
        0&\Meta\left(\Delta^{\mu\nu}\delta_{ab}+e^\mu_be^\nu_a-\frac23e^\mu_ae^\nu_b\right)+\Mzeta e^\mu_ae^\nu_b\\
    \end{pmatrix}
    \,.
\end{equation}

The leading term in the equations of motion \eqref{eq:u.dW2},
\begin{equation}
    u\cdot\cfd W_{a_1a_2}^{(\bm q_1,\bm q_2)} = 2\left[iq_1\cdot A_{a_1}^b W_{ba_2}^{(\bm q_1,\bm q_2)}+\dots\right]_{\overline{12}}\,,
\label{eq:u.dW2=qAW}\end{equation}
drives (complex) oscillations of the Wigner function. The frequency of these oscillations is of order $q$, which is much higher than the relaxation rate of the fluctuations $\mathcal O(q^2)$. One can take advantage of this separation of scales by averaging over faster oscillations (similarly to the rotating wave approximation in optics). To achieve this we need to consider the normal modes of the oscillations driven by matrix $q\cdot A$.

The frequencies and the normal modes are determined by the eigenvalues and eigenvectors of matrix $q\cdot A$:
\begin{equation}
  \label{eq:psi}
  q\cdot A_a^{~b}\Psi^A_b(q) = \lambda^{(A)}(q)\Psi^A_a(q) \,.
\end{equation}
There is no summation over index $A$, which takes values from the set
$\{+,-,1,2\}$. Using matrix $A_{~a}^{\mu~b}$ from Eq.~\eqref{eq:AB-linear} we find
\begin{equation}
  \label{eq:Psi-expl}
  \Psi_a^\pm(q) =
  \begin{pmatrix}
   \mp\hat q_a\\  c_s^{-1}
  \end{pmatrix}\,,
  \quad \lambda^{(\pm)}=\pm c_s|q|\,;\qquad
    \Psi_a^{1,2}(q) =
  \begin{pmatrix}
   t^{1,2}_a(q)\\ 0
  \end{pmatrix}\,,\quad
  \lambda^{(1,2)}=0\,.
\end{equation}
where $\hat q_a\equiv q_a/|q|$ is a unit vector. The $\pm$ modes mix longitudinal velocity and energy density (or pressure) fluctuations and thus correspond to sound, which will be our focus here. The 1,2 modes correspond to velocity shear (transverse momentum) modes, $q^a t_a^{1,2}=0$, and will not be needed here.

The dual set of the {\em left} eigenvectors, $\tilde\Psi_A^a(q\cdot A)_a^{~b}=\lambda_{(A)}\Psi_A^b$ (with the same eigenvalues $\lambda_{(A)}=\lambda^{(A)}$), is given
by
\begin{equation}
  \label{eq:Psi-dual}
  \tilde\Psi^a_\pm (q)=
  \frac12
  \begin{pmatrix}
    \mp\hat q_a\\ c_s
  \end{pmatrix}\,;\qquad
    \tilde\Psi^a_{1,2}(q) =
  \begin{pmatrix}
   t^{1,2}_a(q)\\ 0
  \end{pmatrix}\,.
\end{equation}
In principle, the tilde is unnecessary since the $\Psi_a^A$ and
$\tilde\Psi^a_A$ are already distinguished by the location of
indices. The dual eigenvectors are normalized so that
\begin{equation}
  \label{eq:theta-norm}
  \tilde\Psi_A^a\Psi_a^B=\delta_A^B\,.
\end{equation}

Using the completeness of the basis of eigenvectors $\Psi$
we can express fluctuation vector $\phi_a$ in terms of the normal modes $\phi_A$:
$\phi_a=\Psi_a^A\phi_A$. Correspondingly, the correlators of the
variables $\phi_a$ can be expressed in terms of the correlators of $\phi_A$:
$W_{a_1\dots a_N}^{(\bm q_1,\dots,\bm
q_N)}(x)
= W_{A_1\dots A_N}^{(\bm q_1,\dots,\bm
q_N)}(x)\Psi^{A_1}_{a_1}(\bm q_1)\dots\Psi^{A_N}_{a_N}(\bm q_N)$.

We can project equation~(\ref{eq:u.dW2=qAW}) onto the normal mode basis by
multiplying with $\tilde\Psi^{a_1}_{A_1}(\bm
q_1)\tilde\Psi^{a_2}_{A_2}(\bm q_2)$:
\begin{equation}
  \label{eq:PsiPsiudW}
   \tilde\Psi^{a_1}_{A_1}(q_1)\tilde\Psi^{a_2}_{A_2}(q_2)\times u\cdot\cfd W_{a_1a_2}^{(q_1,q_2)}
= u\cdot\cfd W_{A_1 A_2}^{(q_1,q_2)} + \mbox{ (terms with gradients of
$\tilde\Psi$)}\,.
\end{equation}
Since $\tilde\Psi$ is the left eigenvector of $q\cdot A$, $\tilde\Psi^a_A(q\cdot A)_a^{~b}=\lambda_{(A)} \tilde\Psi^a_A$, the leading, $\mathcal O(q)$, term in Eq.~\eqref{eq:u.dW2=qAW} is diagonalized:
\begin{equation}
  \label{eq:qAW}
  \tilde\Psi^{a_1}_{A_1}\tilde\Psi^{a_2}_{A_2}\times 2\left[
  iq_1\cdot A_{a_1}^{~b}
    W_{ba_2}\right]_{\overline{12}} =
    2i\left[\lambda_{(A_1)}( q_1)\right]_{\overline{12}}  W_{A_1A_2}\,.
\end{equation}
Thus, the oscillating frequency of each normal mode pair correlator $W_{A_1A_2}$ is given by the sum of the corresponding eigenvalues: $2\left[\lambda_{(A_1)}( q_1)\right]_{\overline{12}}=
\lambda_{(A_1)}( q_1)+\lambda_{(A_2)}( q_2)$ \,.
The modes with nonzero $\mathcal O(q)$ frequencies $\pm2c_s|q|$ could be averaged out while describing evolution of modes whose evolution rate is $\mathcal O(q^2)$, or $\mathcal O(k)$. We shall focus on the $-+$ mode which has zero oscillating frequency, since $\lambda_-+\lambda_+=0$. This corresponds to the correlator of sound fluctuations. 

Since, for $N=2$, only one wave-vector is independent: $q_1=-q_2\equiv q$,
we define this projection as follows:
\begin{equation}
    W_{-+}(q)\equiv
    \tilde\Psi_-^{a_1}(q)\tilde\Psi_+^{a_2}(-q)
    W_{a_1a_2}(q,-q)
    =
    \tilde\Psi_-^{a_1}(q)\tilde\Psi_-^{a_2}(q) W_{a_1a_2}(q,-q)\,,
\end{equation}
where we used $\Psi^+(-q)=\Psi^-(q)$, as is evident from Eq.~\eqref{eq:Psi-dual}. Using mutual orthonormality of $\Psi$ and $\tilde\Psi$ sets, we can obtain the coefficients in the equation for $W_{-+}$ by projecting Eq.~\eqref{eq:u.dW2} and using Eqs. \eqref{eq:AB-linear}, \eqref{eq:Psi-expl}, and \eqref{eq:Psi-dual} to calculate quantities such as:
\begin{equation}
     \tilde\Psi^a_-(q) A_{~a}^{\mu~b} \Psi^-_b(q) = -
  c_s \hat q^\mu\,;
 \quad
  \tilde\Psi^a_-(q) \cfd_\lambda A_{~a}^{\mu~b} \Psi^-_b(q) = -
  \partial_\lambda c_s \hat q^\mu\,, \quad\mbox{etc.}
\label{eq:PsiAPsi}
\end{equation}

The resulting equation has the following form:
\begin{equation}
    \mathcal L[W_{-+}(q)]=-\gamma_Lq^2\left(W_{-+}(q)-\frac{w}\beta\right) + \Bt\,W_{-+} \,,
\label{eq:LK}
\end{equation}
where operator
\begin{equation}
    \mathcal L[W_{-+}]\equiv
     (u+c_s\hat q)\cdot\cfd W_{-+} + F_\lambda^- \frac{\partial}{\partial q_\lambda} W_{-+}\,,
\end{equation}
with 
\begin{equation}\label{eq:F-}
    F_\lambda^- = \tilde\Psi^a_-(q) (q_\mu K^\mu_\lambda)_a^{~b} \Psi^-_b(q)= - \partial_\lambda c_s|q| -
    q^\mu\partial_\lambda u_\mu - a_\lambda c_s|q| - 2c_s^2q^\nu\omega_{\nu\lambda}
  \,,
\end{equation}
is identical to the Liouville operator for the kinetic theory of particles with momentum $\bm q$ and energy $E=c_s(x)|\bm q|$. In other words, the particles are phonons with space-time dependent propagation speed $c_s(x)$ as measured in the local rest frame of the fluid. This Liouville operator is derived in Ref.~\cite{An:2019rhf} directly using phonon equations of motion (without using hydrodynamics). Within the formalism we use here it is obtained by substituting $A\to-c_s q$ into 
$q_\mu\KK^\mu_\lambda$ 
given by Eq.~\eqref{eq:K4},
in accordance with Eq.~\eqref{eq:PsiAPsi}. The four terms in Eq.~\eqref{eq:F-} accurately describe all relativistic inertial effects experienced by a phonon due to the inhomogeneity, expansion, acceleration and rotation of the fluid, respectively:
\begin{equation}
    F_\lambda^- = - \partial_\lambda E -
    q^\mu\partial_\lambda u_\mu - E a_\lambda  - 2Ev^\nu\omega_{\nu\lambda}\,,
\end{equation}
where $v^\nu=\partial E/\partial q_\nu = c_s\hat q^\nu$ is the phonon group velocity in the local rest frame of the fluid ($\hat q\equiv q/|q|$).

On the right-hand side of Eq.~\eqref{eq:LK}, the first term describes relaxation of $W_{-+}$ to thermodynamic equilibrium $\bar W_{-+}=w/\beta$ at the well-known (longitudinal) relaxation (sound attenuation) rate
\begin{equation}
    \gamma_L\equiv \frac1 w\left(\frac43\eta+\zeta\right)\,.
\end{equation}

The last term in Eq.~\eqref{eq:LK} is proportional to a linear combination of average fluid velocity gradients
\begin{equation}
    \Bt\equiv -\hat q^\mu\hat q^\nu\partial_\mu u_\nu -(1+c_s^2+\dot c_s)\,\partial\cdot u - c_s^{-1}(1+2c_s^2)\hat q\cdot a\,,
\end{equation}
where $\dot c_s$ is the logarithmic derivative of $c_s$ with respect to entropy: $\dot c_s\equiv d\ln c_s/d\ln s$, as defined in Ref.~\cite{An:2019rhf}. It comes from the $\tilde\Psi_-\Psi^-$ projection in Eq.~\eqref{eq:PsiAPsi} of matrices $B$ and $\Delta B\equiv(1-1/N)\KK_\lambda^\lambda$ ($N=2$) given by Eqs.~\eqref{eq:AB-linear} and \eqref{eq:DeltaA}, as well as from the $\tilde\Psi_-$ gradient terms in Eq.~\eqref{eq:PsiPsiudW}.
While it is already remarkable that the hydrodynamics reproduces the Liouville operator with all the inertial effects, there is something even more remarkable. The velocity gradient term in Eq.~\eqref{eq:LK} can be written as
\begin{equation}
    \Bt= \frac{\mathcal L[wc_s|q|]}{wc_s|q|}\,.
\end{equation}
This means that a rescaled Wigner function 
\begin{equation}
    N_{-+}\equiv\frac{W_{-+}}{wc_s|q|}
\end{equation}
obeys an equation where the velocity gradient term $\Bt$ is completely cancelled:
\begin{equation}
    \mathcal L[N_{-+}]=-\gamma_Lq^2\left(N_{-+}-\frac{1}{\beta c_s|q|}\right)\,.
\end{equation}
What is also remarkable is that this rescaled Wigner function has equilibrium value, to which it relaxes, equal to $T/E$ -- the classical (Rayleigh-Jeans) limit of the Bose-Einstein distribution of particles with energy $E=c_s|q|$ at temperature $T=1/\beta$, as the phonon phase space distribution would do.

\section{Summary and conclusions}
\label{sec:conclusion}

In this paper we introduced a novel formalism for describing fluctuations in {\em relativistic\/} hydrodynamics. The focus of this formalism is on {\em non-Gaussianity} of fluctuations, which is of primary interest in the search for the QCD critical point in heavy-ion collisions. Below we briefly outline the main ingredients of the formalism we introduced.

\begin{itemize}
    \item 

In the deterministic approach, fluctuations are described by equal-time correlation function of hydrodynamic variables. {\em Relativistic} flow presents interesting challenges. In particular, what is the frame in which ``equal time'' is being defined? Our formalism addresses this challenge by introducing averaged rest frame of the fluid whose four-velocity $u(x)$ satisfies Landau conditions for {\em averaged\/} energy-momentum tensor given by Eq.~\eqref{eq:ave-uT}. 

  \item 
Using the field $u(x)$ we define connection such that transporting vector $u(x)$ from point $x$ to point $x'$ results in vector $u(x')$. Using this connection we define {\em confluent derivative} which is a derivative under which the four-vector $u(x)$ behaves as a constant --- Eq.~\eqref{eq:barnabla-def}.

  \item 
At each space-time point we then introduce a local Euclidean basis triad $e_a^\mu(x)$, orthogonal to four-vector $u(x)$ --- Eq.~\eqref{eq:e.u}. We extend the definition of the confluent derivative is such a way that the basis vectors are also confluently constant --- Eq.~\eqref{eq:nabla-e}.

There is a freedom of choice of the local basis, which represents a redundancy of description, since physics does not depend on this choice. The corresponding SO(3) local gauge invariance can be used to our advantage by making the formalism manifestly gauge covariant.

  \item 
We define the fluctuating hydrodynamic variables as conserved
densities measured in the averaged rest frame of the fluid --- Eqs.~\eqref{eq:uT-e-pi} and~\eqref{eq:uJ}. The three independent momentum density variables $\pi$ can be expressed using the local Euclidean triad basis.

  \item 
To describe fluctuations, we define the $N$-point {\em confluent} correlator $\barG_{(N)}$ --- Eq.~\eqref{eq:H}, which  (a) is an equal-time correlator in the frame defined by the four-velocity $u(x)$ at the midpoint $x$ --- Eq.~\eqref{eq:x}, and (b) is SO(3) covariant, i.e., transforms as a local tensor at the midpoint --- Eq.~\eqref{eq:H-TH}.

  \item 
We define confluent derivative of $\barG_{(N)}$, which is also SO(3) covariant --- Eq.~\eqref{eq:epsnabla}.

  \item 
Following Ref.~\cite{An_2021}, we define generalized Wigner transform $W_{(N)}$ --- the $3N$-fold Fourier integral of $\barG_{(N)}$ with the midpoint fixed --- Eq.~\eqref{eq:WG}. This allows us to take advantage of hydrodynamic scale separation $k\sim q^2\ll q$.

\end{itemize}

The evolution of Wigner functions  $W_{(N)}$ are given by Eqs.~\eqref{eq:W2-eqs} for $N=2$ and \eqref{eq:W3-eqs} for $N=3$.
The evolution equations have diagrammatic representation given in Fig.\ref{fig:gamma-diagrams}.

These equations simplify if the variables are chosen to be conserved densities, such as $\pi$, $\teps$, $\tn$ defined in Eqs.~\eqref{eq:uT-e-pi} and \eqref{eq:uJ}. 
The simplifications are described in Section~\ref{sec:evolution-conserved} and the resulting equations are Eqs.~\eqref{eq:W2-cons} and \eqref{eq:W3-cons}. Equation \eqref{eq:W3-cons}, or more generally, Eq.~\eqref{eq:W3-eqs}, for the evolution of the $N=3$ correlator is the most important new result of this work.

In the context of the optimization of the variable choice, it is worth keeping in mind that connecting hydrodynamic fluctuations to particle multiplicity fluctuations measured in experiments requires a procedure, known as freezeout, of converting fluctuations of hydrodynamic variables to the fluctuations of particles. The main requirement for such a procedure is the adherence to conservation laws. It is thus more natural to consider fluctuations of the conserved densities in terms of which the implementation of the conservation laws is straightforward. In fact, in the maximum entropy approach to freezeout introduced in Ref.~\cite{Pradeep:2022eil}  the ingredients needed to implement the constraints due to the conservation laws are correlation functions of conserved densities.

The operator $\VABDW$ driving the evolution of the Wigner functions has a very intuitive interpretation in the case of $N=2$ equation upon projection on the normal modes of the ideal hydrodynamics. The components of $W$ corresponding to the sound-sound channel obey the same equation as the phonon phase-space distribution function as discussed in Section~\ref{sec:phonon}.

As discussed in Section~\ref{sec:evolution-conserved}, the choice of conserved densities as variables has yet another advantage. It is interesting and aesthetically pleasing that in the equation for $N=3$ correlator, apart from the dislocation term, the coefficients are simply derivatives of the coefficients of the $N=2$ equations.

While we focused on $N=3$ correlators of hydrodynamic fluctuations, the approach we introduce in this paper should be applicable to $N=4$ correlators. Given the somewhat intuitive form of the equation we derived, their generalization to $N=4$ case could be partially guessed, but we shall defer a more careful derivation to future work.

It should be also interesting to use the formalism introduced here to study the nonlinear feedback of non-Gausian fluctuations known as long-time tails in correlation functions, as it is done, e.g., in Refs.~\cite{Andreev:1978,Akamatsu:2017,Akamatsu:2018,Martinez:2018,An:2019rhf,An:2019fdc} for Gaussian fluctuations.

Finally, the search for the critical point in heavy-ion collision experiments requires quantitative understanding of the dynamics of the fluctuations, in particular, non-Gaussian fluctuations. The ultimate goal of this work is to achieve this necessary understanding. In order to apply the equations we derived to realistic scenarios of heavy-ion fireball expansion, numerical implementation is required. This is not an easy task, given the many variables and components of the correlators involved. More work is needed to develop efficient methods of simulating these evolution equations.

\begin{acknowledgments}
This work is supported by the European Research Council (ERC) under the European Union’s Horizon 2020 research and innovation programme (grant number: 101089093 / project acronym:
High-TheQ) (X.A.), the National Science Foundation CAREER Award PHY-2143149 (G.B.), and the U.S. Department of Energy, Office of Science, Office of Nuclear Physics Grant No. DEFG0201ER41195 (M.S.).
\end{acknowledgments}

\appendix

\section{Notations}
\label{sec:notations}

Below are some of the notations used throughout this paper to help the reader look up their definitions.

\begin{list}{}{}    

\item $\big[\dots\big]_\oN$ --- average over permutations of indices $1\dots N$, e.g., $\big[X_{12}\big]_{\overline{12}}=\frac12\big(X_{12}+X_{21}\big)$.

\item $\cfd$ --- confluent derivative, like a covariant derivative, depends on the type of the object it is acting upon. For fields --- Eqs.~\eqref{eq:barnabla-def},~\eqref{eq:nabla-pi},~\eqref{eq:nabla-e},~\eqref{eq:Dphi-Delta},~\eqref{eq:Dphi-omega}, for a confluent correlator --- Eq.~\eqref{eq:epsnabla}.

\item $\dxone$ --- derivative of the correlator w.r.t. one of the endpoints --- Eq.~\eqref{eq:nablax1H-nablaphi}.

\item $\bdyone$ ---  balanced derivative w.r.t. one of the separation vectors of a correlator --- Eqs.~\eqref{eq:dy1}.

\item $\dyone{\lambda}$  --- balanced derivative represented as a four-vector --- Eq.~\eqref{eq:dy-bdy}.

\item $\Delta_h$ --- finite difference, Eq.~\eqref{eq:Deps-def}. Usually, $h\to0$.

\item $\bm\Omega$ --- Eq.~\eqref{eq:Omega-def}.

\item $\alpha$ --- the variable conjugate to charge density $\tilde n$ (and $\breve n$), i.e., $\mu/T$ --- Eq.~\eqref{eq:d-tilde-s}.

\item $\tilde\beta$ --- the variable conjugate to $\teps$ (if $\bm v=0$ --- same as inverse temperature) --- Eq.~\eqref{eq:d-tilde-s}.

\item $\breve\beta$ --- the variable conjugate to $\breve\varepsilon$, i.e., inverse temperature --- Eq.~\eqref{eq:beta-beta}.

\item $\varepsilon$ --- average energy density in average Landau frame, Eq.~\eqref{eq:ave-uT}.

\item $\teps$ --- fluctuating energy density in the average Landau frame $u$ --- Eq.~\eqref{eq:uT-e-pi}

\item $\breve\varepsilon$ --- fluctuating energy density in the fluctuating Landau frame $\breve u$ --- Eq.~\eqref{eq:breve-uT}.

\item $\eta^{\widehat c}$ --- noise field --- Eq.~\eqref{eq:eta-eta}.

\item $\pi_\nu$ --- fluctuating momentum density in the average Landau frame, Eq.~\eqref{eq:uT-e-pi}.

\item $\bm\pi$ --- three-vector representing $\pi_\nu$ in the basis $\bm e$: $\bm\pi=\bm e^\nu\pi_\nu$.

\item $\phi_a$ --- fluctuation of $\tpsi_a$ --- Eq.~\eqref{eq:phi}.

\item $\chi^a$ --- variable conjugate to $\psi_a$ --- Eqs.~\eqref{eq:ds},~\eqref{eq:chi-beta}. 

\item $\psi_a$ --- average value of $\tilde\psi_a$ --- Eq.~\eqref{eq:phi},
e.g., $\langle\{\bm \pi, \teps, \tn\}\rangle=\{\bm 0,\varepsilon,n\}$.

\item $\tpsi_a$ --- fluctuating hydrodynamic fields, e.g., $\{\bm \pi, \teps, \tn\}$.

\item $\ucon$ --- confluent connection, Eqs.~\eqref{eq:barnabla-def},~\eqref{eq:bar-omega}.

\item $\mathring\omega$ --- SO(3) connection --- Eq.~\eqref{eq:nabla-e},~\eqref{eq:ring-o}, generator of $R$, Eq.~\eqref{eq:R-omega}.

\item $A^{\mu~b}_{~a},\,B_a,\,C^{\mu}_{\widehat c  a},\,\Ct^{\mu~b}_{\widehat c a},\,D^{\mu\nu~b}_{~~a},\,\Dt^{\mu\nu~bc}_{~~a}$ --- coefficients in Eq.~\eqref{eq:u.nablapsi-ABCD}.

\item $A^\mu_{(3)}$, $B_{(3)}$, $C^\mu_{(3)}$, $\Ct^\mu_{(3)}$, $D^{\mu\nu}_{(3)}$, $\Dt^{\mu\nu}_{(3)}$ --- order-3 hypermatrices (3 suppressed indices) --- Eq.~\eqref{eq:u.phi-ABCD}.

\item $E^{\mu\nu}_{ab}$ --- Eq.~\eqref{eq:DWQ}.

\item $\barG_{(N)}$ --- confluent correlator with $N$ indices and arguments suppressed --- Eqs.~\eqref{eq:H}, \eqref{eq:G-sup}, \eqref{eq:GN}.

\item $K^\mu_\lambda$  --- Eqs.~\eqref{eq:DeltaA},~\eqref{eq:K4}.

\item $N$ --- the number of correlated fields, i.e., the order of a correlator, e.g., Eq.~\eqref{eq:rawG} or Eq.~\eqref{eq:H}. 

\item $Q^{\mu\nu}_{ab}$ --- the coefficient of the noise-induced term in deterministic equation --- Eq.~\eqref{eq:Q}.

\item $Q^{\mu\nu}_{(3)}$, $\Qt^{\mu\nu}_{(3)}$ --- order-3 hypermatrices in the noise-induced terms --- Eq.~\eqref{eq:Q3C'C}.

\item $R(x,x')$ --- finite SO(3) connection --- Eqs.~\eqref{eq:ReLe}, related to $\mathring\omega$ by \eqref{eq:R-omega}.

\item $W_{(N)}$ --- Wigner transform of $\barGN$ --- Eq.~\eqref{eq:WG}.

\item $\aoA_\lambda$ --- Eq.~\eqref{eq:ao}.

\item $a,\,b,\,\dots$ --- first few lowercase Latin letters refer to hydrodynamic fields $\psi_a,\,\psi_b,\,\dots$. They take values from the set $\{1,2,3,\dots\}$, referring to the three components of momentum density, followed by scalar variables, such as energy and charge densities.

\item $(\dots)^{,a}$ --- derivative with respect to $\psi_a$ --- Eqs.~\eqref{eq:A3A'}, \eqref{eq:ds}.

\item $(\dots)_{,a}$ --- derivative with respect to $\chi^a$ --- Eq.~\eqref{eq:ja-ideal}.

\item $e_a^\mu(x)$ or 
$\bm e(x)$ --- orthogonal triad of four-vectors at point $x$ --- Eq.~\eqref{eq:e.u}. $e_a\neq 0$ only for $a=1,2,3$.

\item $\Dx$ --- four-vector displacement, usually, infinitesimal.

\item $\Dxt$ --- infinitesimal scalar, displacement in $u$ direction --- Eq.~\eqref{eq:Dxt}.

\item $\tn$, $\breve n$ --- similar to $\teps$ and $\breve\varepsilon$, but for charge density instead of energy density --- Eq.~\eqref{eq:uJ}.

\item $p$ --- pressure --- Eq.~\eqref{eq:pressure}.

\item $\bm q_1$ or $q_{1\mu}$ --- spatial wave vector conjugate to $\bm y_1$ --- Eq.~\eqref{eq:WG}. $q_{1\mu}\equiv\bm e_{\mu}\cdot\bm q_1$.

\item $\tilde s$, $\breve s$ --- similar to $\teps$ and $\breve\varepsilon$, but for entropy density instead of energy density.

\item $u$ --- four-velocity of the average Landau frame --- Eq.~\eqref{eq:ave-uT}.

\item $\bm v$ or $v^\mu$ --- fluctuating fluid velocity in the average Landau frame --- Eqs.~\eqref{eq:v-beta}, \eqref{eq:v}.

\item $x$ --- usually refers to the midpoint of a correlator, Eq.~\eqref{eq:x}.

\item $x_1$ --- refers to one of the endpoints of a correlator --- Eq.~\eqref{eq:rawG},~\eqref{eq:H}.

\item $\bm y_1$ or $y_1$ --- refers to one of the separation vectors of a correlator --- Eq.~\eqref{eq:xxy}.

\end{list}

\bibliographystyle{utphys}
\bibliography{references}

\end{document}